\documentclass[preprint,amsmath,amssymb,superscriptaddress,showpacs,pre]{revtex4-1}

\usepackage{graphicx}
\usepackage{dcolumn}
\usepackage{bm}
\usepackage{color}
\usepackage{epstopdf} 
\usepackage{amsmath,amsfonts,amssymb,amsthm}
\usepackage{algorithm}
\usepackage{algorithmic}
\usepackage{braket}
\usepackage{mathbbol}
\usepackage{subfigure}

\usepackage{lineno}
\usepackage{nicefrac}       
\usepackage{microtype}      

\let\originalleft\left
\let\originalright\right
\renewcommand{\left}{\mathopen{}\mathclose\bgroup\originalleft}
\renewcommand{\right}{\aftergroup\egroup\originalright}

\def\<{\langle}
\def\>{\rangle}
\def\vx{\vec x}

\def\vxh{\vec{x}^{\:h}}

\def\vXh{\vec{A}}
\def\vXd{\vec{D}}

\def\Imat{\mathbb{I}}

\DeclareMathOperator{\Tr}{Tr}

\newcommand{\ddroit}{\textrm{d}}

\newcommand{\Lam}{\bm{\Lambda}}
\newcommand{\Sig}{\bm{\Sigma}}

\begin{document}

\title{Ancestral Sequence Reconstruction for Co-evolutionary models}
\date{}

\author{Edwin Rodr\'{i}guez Horta} 
\email{Correspondence to: Edwin Rodr\'{i}guez Horta, \bf{edwin@fisica.uh.cu}}\affiliation{Sorbonne Universit\'{e}, CNRS, Institut de Biologie Paris-Seine, Laboratoire de Biologie Computationnelle et
Quantitative -- LCQB, Paris, France}
\affiliation{Group of Complex Systems and Statistical Physics, Department of Theoretical Physics, University of Havana, CP10400. La Habana, Cuba}

\author{Alejandro Lage-Castellanos} 
\affiliation{Group of Complex Systems and Statistical Physics, Department of Theoretical Physics, University of Havana,  CP10400. La Habana, Cuba}

\author{Roberto Mulet} 
\email{Correspondence to: Roberto Mulet, \bf{mulet@fisica.uh.cu}}
\affiliation{Group of Complex Systems and Statistical Physics, Department of Theoretical Physics, University of Havana,  CP10400. La Habana, Cuba}

\begin{abstract}

   The ancestral sequence reconstruction problem is the inference, back in time, of the properties of common sequence ancestors from  measured properties of contemporary populations. Standard algorithms for this problem  assume independent (factorized)  evolution of the characters of the sequences, which is generally wrong (e.g. proteins and genome sequences). In this work, we have studied this problem for sequences described by global co-evolutionary models, which reproduce the global pattern of cooperative  interactions between the elements that compose it. For this, we first modeled the temporal evolution  of correlated real valued characters  by a multivariate Ornstein-Uhlenbeck process on a finite tree. This  represents sequences as Gaussian vectors evolving in a quadratic potential, who describe the selection forces acting on the evolving entities.  Under a Bayesian framework, we  developed a reconstruction algorithm for these sequences  and obtained an analytical expression to quantify the quality of our estimation. We extend this formalism to discrete valued sequences by applying our method to a Potts model. We showed that for both continuous and discrete configurations,  there is a wide range of parameters where, to properly reconstruct the ancestral sequences, intra-species correlations must be taken into account. We also demonstrated that, for sequences with discrete elements, our reconstruction algorithm outperforms traditional schemes based on independent site approximations.

\end{abstract}

\maketitle

\section{Introduction}
\label{sec:int}

The ancestral reconstruction (AR)  problem is the inference, back in time, of the properties of common ancestors using as data set the measured properties of contemporary populations \cite{AncestralReconstruction}. Ancestral reconstruction  rests in a phylogeny, a tree that orders the populations. The leaves of the tree form the contemporary populations (the observed elements). They are connected to common ancestors at branching points, or nodes. The goal of AR is to estimate the internal characteristics of these nodes. Notice that although the construction of this tree (the phylogeny) is a problem itself \cite{Felsenstein}, for many applications of AR this tree is assumed as known, and we follow this approach here.

Ancestral reconstruction relies also on a model of evolution. But, since the actual evolutionary process is rarely known, the proper selection of the model is fundamental for the outcome. In general, the use of simple models deteriorates the inference faster with increasing evolutionary time, but more realistic models are more difficult to calculate. It is in the researcher's ability to properly fine tune the complexity of the model for a specific application.

Ancestral sequence reconstruction (ASR) (\cite{Yang95,Koshi96}) is a sub-problem of AR in which the characteristics of the system under study are described by a sequence of elements. These elements are encoded by character states (residues). In biological applications, these characters are usually defined within a finite set, such as nucleotides for genome sequences or amino-acids for proteins. In this case, the main goal is to infer each of these characters for each of the sequences that conform to the ancient populations.

Modern methods for ASR are based on the Maximum Likelihood (ML) framework \cite{pagel99,pupko}. In this context the elements of the sequences are chosen,  given the  model of evolution and a phylogenetic tree, maximizing the probability of occurrence of the data (sequences) at the bottom of the tree. However, even in systems of moderate sizes, the huge number of possible configurations makes the problem intractable in general and one must introduce extra assumptions to approach real problems. A widely used hypothesis considers the evolution of the sequences as single site independent processes~\cite{Yang95}. This simplification reduces the computational cost of the inference, and is at the basis of every advanced algorithm currently in use (\cite{Yang07,pupko}). These algorithms may vary by type of biological information, i.e. the evolutionary model and the tree used,  and on the strategy to reach the globally optimal solution of the problem. For example, marginal and joint reconstruction or empirical and hierarchical Bayesian method \cite{Huelsenbeck01}. The assumption of single site evolution, however, has been unavoidable, and its effect on the ASR remains unclear. 

We know  that in many biological scenarios this assumption is not correct. For amino acids sequences of proteins there is an abundant evidence of  epistasis \cite{Breen,Harms}, due to structural constrains imposed by the three-dimensional fold of the protein \cite{Olson,Rollins}. In the case of genome  sequences, epistasis is reflected in the distribution over genotypes in a population evolving with sufficient amount of exchange of genetic material (recombination, or any form of sex), a phenomenon called  Quasi-Linkage Equilibrium and discovered by M. Kimura \cite{Kimura,Gao2019}. Then, it is already well accepted that global co-evolutionary models are necessary to correctly represent  relevant statistical features of biological sequences (\cite{Nguyen17,Levy17,MW18}). They have  been fundamental in the   prediction of non-trivial structural contacts in the protein fold \cite{Weigt11},  designing novel functional protein
sequences\cite{prot_design} and the inference of gene interactions \cite{HL}.

Of course the use of the single-site approximation is not exclusive of the ASR methods. It has been identified as a limitation in other important inference  problems in biology. This simplification has been shown to decrease the accuracy of the inferred phylogenetic trees in the presence of non-independent sites \cite{Huelsenbeck99,Nasrallah2011}. However, data sets with strong functional or structural constraints are often analyzed within phylogenetic frameworks that assume independence among sites. Moreover, one of the  most important computational problems in biology: the construction of sequence alignments is typically addressed through profile models, which capture position specificities like conservation in sequences but assume an independent evolution of different positions. A recent attempt to  overcome the limitations of profile models and to include co-evolution among positions was carried out by Muntoni et al in \cite{Muntoni2020}. The search for fast statistical tools, able to improve over the single site approximation is a research line that is still in its infancy.

In this contribution, we assess the impact of intra-species traits correlations on the performance of ancestral reconstruction.  We first study the ASR problem for sequences whose elements are continuous characters co-evolving through an Ornstein-Uhlenbeck (OU) process on a  phylogenetic tree. Although, at first, this model may seem distant from realistic biological applications, it has been used in the field of phylogenetic comparative methods (PCM) \cite{bartoszek_phylogenetic_2012,mitov_fast_2020} and, in a recent contribution \cite{erh2021}, to extract intrinsic signals from hierarchically correlated data. At the same time it opens the way for algorithmic solutions as well as analytical calculations clarifying the relevance of the parameters of the model and the affordability of the technique. With this understanding, we extend the approach to the more biologically relevant case of sequences defined on discrete values. We will show that the main picture devised for the OU process continues to be valid, and also that our reconstruction process outperforms standard methods that assume single site evolutionary processes. Notice that, as far as we know,  a study about the robustness of  current ASR methods to the violation of the hypothesis of independent substitutions  has not been carried out. However a similar study exists for the  phylogenetic inference problem \cite{Nasrallah2011} where the authors found that, for all the methods studied, even small amounts of dependencies can lead to significant errors in estimating the actual topologies. Then, it is very plausible that ASR methods present similar robustness problem. As we will see below our results point into this direction.

The rest of the document is organized as follows.  In section \ref{sec:prob-def} we define mathematically the problem. Then, in \ref{sec:multi-oudyn} we show how this formalism translates into a Bayesian framework  assuming a co-evolutionary model with an Ornstein-Uhlenbeck dynamics. This section includes  an analytical expression for the error in the inference that successfully compares with results from numerical simulations. Next we show how to exploit the mapping from discrete to continuous variables proposed in \cite{Baldassi} to use the results from the previous sections to recover sequences on a finite alphabet subject to a stochastic dynamics. Finally we present the conclusions of our work.

\section{Statement of the problem } 
\label{sec:prob-def}

Let us consider a set of observed sequences $ \left\lbrace \underline{X}^m\right\rbrace_{m=1,\dots,M} $  phylogenetically related by an evolutionary process on a tree $\mathcal{T}$ that we will always assume known. Each sequence $\underline{X}=(X_1,\dots,X_L)$ has length $L$ and  character states $X_i$ take values on a discrete alphabet $\boldsymbol{X}$ with size $q=|\boldsymbol{X}|$. For genome or protein sequences, the alphabet would be that of the 4  bases of DNA or the 20 amino acids respectively. We assume the sequences to be aligned, conforming a matrix of dimensions $M\times L$ called multiple sequences alignment (MSA).

These contemporary  sequences correspond to the terminal nodes of the tree (see Fig. \ref{fig:treeXX}) and are assumed to be the result of an evolutionary process initialized from a common ancestor at root $\underline{X}^{\:0}$. The evolution is  mathematically defined by a  propagator $P(\underline{X}^j|\underline{X}^i,\Delta t_{ij})$ representing the probability of observing sequence $\underline{X}^j$ as the result of the evolution from $\underline{X}^i$ in a time  $\Delta t_{ij}$. The precise form of the evolutionary propagator, as well as the inference of the phylogenetic tree are problems 
themselves, but to the purpose of this research, they are considered known. 

Our goal is to infer the set of ancestral sequences at the internal nodes of the tree $\left\lbrace \underline{A}^m\right\rbrace$ from which the observed contemporary sequences $\left\lbrace \underline{X}^m\right\rbrace$ evolved.   Bayesian methods  use to  compute the maximum-a-posteriori (MAP) estimate,  by maximizing the joint probability of ancestral configurations given the contemporary sequences and the details of the evolution (tree and  propagator), which could be evaluated via Bayes rule :

\begin{equation}
\label{bayes_theo}
P(\left\lbrace \underline{A}^m\right\rbrace |\left\lbrace \underline{X}^n\right\rbrace)\propto P(\left\lbrace \underline{X}^n\right\rbrace|\left\lbrace \underline{A}^m\right\rbrace)*P(\left\lbrace \underline{A}^m\right\rbrace )
\end{equation}

Furthermore, it is common to assume an uninformative prior distribution for sequences $\left\lbrace \underline{A}^m\right\rbrace$ by taking $P(\left\lbrace \underline{A}^m\right\rbrace)$ uniform,  and the inference becomes  the calculation of the   \textit{Maximum Likelihood Estimate} (MLE) given by

\begin{equation}
\{\underline{A}^m\}^*= \max_{\left\lbrace \underline{A}^m\right\rbrace} \left\lbrace P(\left\lbrace \underline{X}^m\right\rbrace|\left\lbrace \underline{A}^m\right\rbrace)\right\rbrace 
\label{likelihood}
\end{equation} 

Most efficient methods to compute the  likelihood  function on a tree are based on a  dynamic  programming  algorithm called Felsenstein's pruning algorithm \cite{Felsenstein}, which exploits the  recursion equation 
\begin{equation}
P^{n}(\left\lbrace \underline{X}^m\right\rbrace| \underline{A}^n)=\prod_{m\in C(n)}\left[ \sum_{\underline{A}^m} P^{m}(\left\lbrace \underline{X}^m\right\rbrace|\underline{A}^m)\: P(\underline{A}^m|\underline{A}^n)\right] 
\label{FPA}
\end{equation} 
where $P^{n}$ represent the  conditional probability of observing all existing data sequences that share node $n$ as an ancestor given that the sequence of this ancestor is $\underline{A}^n$. The term $C(n)$ denotes all children nodes of node $n$, and $P(\underline{A}^m|\underline{A}^n)$ is the evolutionary propagator. Then, to compute the  likelihood of the observed data, as a function of the sequence configuration at each internal node, the expression (\ref{FPA}) is evaluated starting from the leaves, where $P^{n}(\underline{A}^n)=\delta_{\underline{A}^n,\underline{X}^n} $, to the root  of the tree.

Since, for systems of realistic sizes, the number of   possible sequences at the internal nodes of the tree is huge,  the evaluation of the likelihood via the recursion relation (\ref{FPA})  is intractable, and therefore the  solution  sought  through an optimization scheme impossible.  To reduce the phase space, the standard assumption is to consider an independent-site approximation, where each site of the sequences evolves independently of all others.   This allows to write equation (\ref{FPA}) for each sequence site, being the observed data at leaves of the tree a single column of the aligned sequences.  Therefore,  $P(\underline{A}^j|\underline{A}^i,\Delta t_{ij})$ is factorized leading to a probabilistic reversible model given  by character states frequencies  $P_a(A_a)$ and propagator $P_a(A^i_a|A^j_a,\Delta t_{ij})$ which describe the replacement of  character in the $a$ position of the sequence at node $j$, $A^j_a$, by a character in the same position in the sequence at node $i$, $A^i_a$,  after an evolutionary time $\Delta t_{ij}$. The  selection of the $P_a(A^i_a|A^j_a,\Delta t_{ij})$ depends on the nature of the problem, and may reflect extra biological information introduced in the model.

However, as we  mentioned in the introduction,  co-evolutionary processes could be relevant both in   proteins families and in genome sequences. In this work we are going to avoid the assumption of independent-site evolution, and use a Bayesian formalism, considering instead that each sequence is better described by a pairwise Potts model :
\begin{equation}
P(\underline{X})=\frac{1}{Z} \exp \left\lbrace \sum_{1\leq i<j\leq L} \lambda_{ij} (X_i,X_j)\right\rbrace 
\label{Potts}
\end{equation} 
where statistical couplings $\lambda_{ij}$ between sites encode the epistatics signal of the system. This model has been widely used in biological systems \cite{Nguyen17}, and it is the least biased statistical model that reproduces the empirical frequencies of characters by site and by site pairs, which are the most common statistical observables for biological data.

\section{Multivariate Ornstein-Uhlenbeck dynamics}
\label{sec:multi-oudyn}

All phylogeny based method assume that a single evolutionary history underlies the sample of the sequences under study, then recombination and gene flow are ignored because these may give rise to graphs that are no longer trees but networks. Therefore the forces of evolution consistent with our framework are those that act on the single genotype level as selection, mutations and genetic drift. In this context  we propose to use a multivariate OU dynamics, which can  take into account selection when the  potential is considered to be a fitness proxy, but not changes in selection rules on time, because this would invalidate the assumption of stationary evolution. However, it may consider changes in the mutation rate when times are measured in terms of a molecular clock rather than in physical time.

Although the discrete nature of biological evolution is an admitted fact since the discovery of the genetic code, there are good reasons to study AR problems in continuous variable models as the multivariate OU dynamics. First, correlation between traits could be present also at the phenotypic level, where characters are real valued quantities, as body mass. Second, even at the gene level, there are some tricks that help to turn discrete into continuous  variables \cite{Baldassi}, thus rendering the continuous approach applicable to the discrete case. And, finally, continuous variables may simplify the problem enough to allow for a precise analytical description, which can illustrate the relevance of the parameters of the problem.

\subsection{Formalism}
\label{ssec:formalism}

We will first study a  model of phylogenetic tree; one in which each specie is described by a continuous vector $\vx \in \mathbb{R}^L $ in $L$ dimensions.  Furthermore, we assume  that the  evolution of these characters follow an OU process (see details in \ref{app:OU}) . For this, continuous degrees of freedom evolves under a potential $V(\vec{x})=\frac{1}{2} \vec{x}^T\bm C ^{-1} \vec{x}$ leading to the stationary distribution:

\begin{equation}
P^0(\vec{x}) \propto \exp-\frac{1}{2}\left\{ \vec{x}^T\bm{C}^{-1}\vec{x} \right\}
\label{eq:gauss_multiv}
\end{equation}

For and OU process like this the corresponding propagator, i.e. the solution of the Fokker-Planck equation, is  given by :
\begin{equation}
\label{eq:propagator_OU}
\begin{split}   
P^0(\vec{x}_j|\vec{x}_i;\Delta t_{ij}) &\propto \exp-\frac{1}{2}\left\{ \vec{x}^T_j\Sig^{-1}_{ij}\vec{x}_j+ \vec{x}^T_i\Lam^2_{ij}\Sig^{-1}_{ij} \vec{x}_i - 2 \vec{x}^T_i \Lam_{ij}\Sig^{-1}_{ij} \vec{x}_j \right\}
\end{split}
\end{equation}
where 
$$ \bm{\Sigma_{ij}} = \bm{C} - \Lam_{ij}\bm{C}\Lam_{ij}, \qquad \Lam_{ij} = e^{-\bm C^{-1} \gamma\Delta t_{ij}}$$

\noindent and where $\bm{C}$ is the correlation matrix, $\Delta t_{ij}$ is the time distance between sequences $\vec{x}_i$ and   $\vec{x}_j$ and $\gamma$ is the characteristic time-scale governing the dynamics.

\begin{figure}[!htb]
	\includegraphics[width=0.7\linewidth]{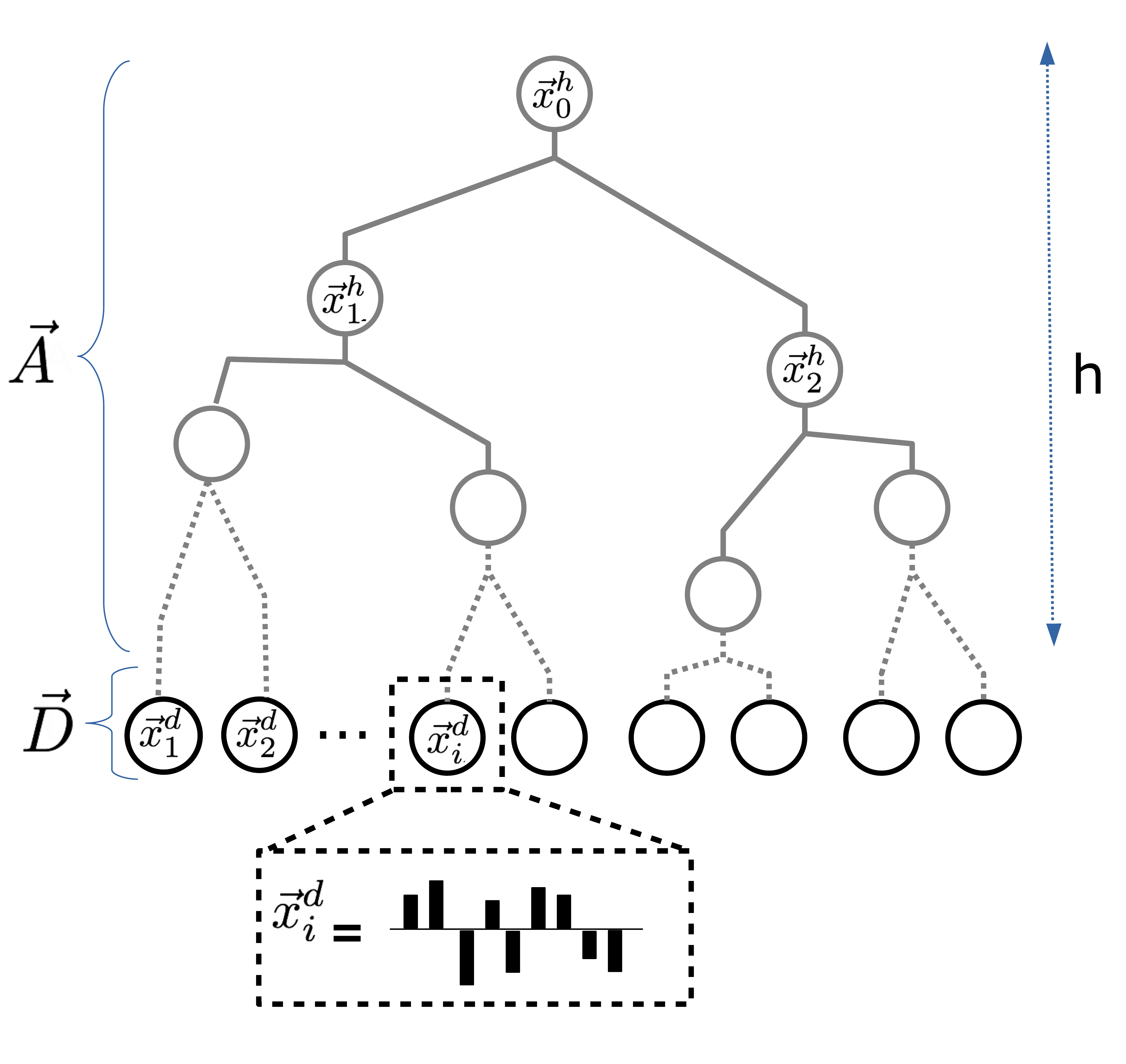}
	\caption{Schematic representation of the evolutionary process in a binary and balanced tree $\mathcal{T}$ with height $h$. The process starts at the root node $0$ with a configuration $\vx_0$ sampled from  $P^{0}(\vx_0)$. The dynamic consists in independent realizations of the OU process on all branches from ancestral nodes to its descendant over times corresponding to the branch length.  The observable data only consist of configurations of the leaf nodes ($\vec{D}$), while configurations of ancestral nodes ($\vec{A}$), remain unknown and must be inferred. }
\label{fig:treeXX} 
\end{figure}

The evolution model is schematized in Figure \ref{fig:treeXX}. It is worth clarifying the conventions used for the notation. We will denote by:

\begin{itemize}
	\item lowercase vector variables (i.e. $\vx,\vx_i,\vx_j,\vxh_1,\ldots$), the real valued vector of dimension $L$ defining the state of the given species.
	\item uppercase vector variables (i.e. $\vec{X},\vXh,\vXd,\ldots$), the concatenation of many state vectors, for instance as the collection of all sequences $\vec{X} = (\vx_1,\ldots,\vx_{N})$.
	\item bold uppercase (i.e. $\bm{\Sigma_{ij}}, \bm{C}, \Lam_{ij}$) $L\times L$ matrices  matrices acting at species level, i.e. over variables $\vx_i$.
	\item  uppercase blackboard bold matrices (i.e. $\mathbb{K},\mathbb{G},\ldots$)   as block matrices at tree level, made of the composition of many species level matrices.
	\item indices $i,j$ to run over the tree nodes, i.e. over the species.
	\item indices $a,b$ to run over  entries of the state vector or concatenated vectors.
\end{itemize}

With these assumptions, the probability of a configuration $\vec{X}$ of the full system (internal nodes configurations and leaves) can be constructed by using  the stationary distribution (\ref{eq:gauss_multiv})  and the propagator (\ref{eq:propagator_OU}). For instance, for the case of a  tree  with binary topology, 
$$ P^0(\vec{X}) = P^0(\vec{x}_0)P^0(\vec{x}_1|\vec{x}_0)P^0(\vec{x}_2|\vec{x}_0)P^0(\vec{x}_3|\vec{x}_1)P^0(\vec{x}_4|\vec{x}_1)\ldots $$

For an arbitrary tree, the probability $P^0(\vec{X})$ can be  shaped as
\begin{equation}
\label{eq:ML_PairwiseDistribution}
P^0(\vec{X}) = \frac{e^{-\mathcal{H}^0(\vec{X})} }{Z^0} 
\end{equation}
invoking the Boltzmann distribution of a system with a pairwise Hamiltonian
\begin{equation}
\label{eq:ML_H0}
\mathcal{H}^0 = -\sum_{i<j}\vec{x}^T_i \bm J_{ij}\vec{x}_j - \sum_{i=1}^N \vec{x}^T_i \bm H_i\vec{x}_i.   
\end{equation}
The interaction terms are $L\times L$ matrices
\begin{equation}
\label{eq:Jij_hom_and_binary}
\bm J_{ij} =
\begin{cases}
\Lam_{ij}\Sig^{-1}_{ij} & \text{if $i$ and $j$ are in contact} ,\\
\bm 0 & \text{otherwise}.
\end{cases}
\end{equation}
and there is a Gaussian local interaction given by
\begin{equation}
\label{eq:Hi_hom_and_binary}
\bm H_i = -\frac{1}{2}\cdot
\begin{cases}
\bm C^{-1}\left[ 1+\sum_{j\in c(0)} \frac{\Lam^2_{0j}}{1-\Lam^2_{0j}}\right] & \text{if $i=0$},\\
\Sig^{-1} _{a(i) i}& \text{if $i$ is a leaf},\\
\bm C^{-1}\left[ \frac{1}{1-\Lam^2_{a(i)i}}+\sum_{j\in c(i)} \frac{\Lam^2_{ij}}{1-\Lam^2_{ij}}\right]& \text{otherwise}.
\end{cases}
\end{equation}
where $c(i)$ and $a(i)$ refers to child and ancestral nodes  of nodes $i$. 

Not surprisingly, the Gaussian nature of the Ornstein-Uhlenbeck survives for the whole tree, and the equilibrium distribution of the concatenated variables $\vec{X}$ is also a Gaussian $P^0(\vec X) \propto \exp(-\frac 1 2 \vec X^T \mathbb{H} \vec X)$, with inverse  covariance matrix
\begin{equation}
 \mathbb{H} = 2 \left(\begin{array}{cccc}
            \bm H_1 & \frac 12\bm J_{12} &\ldots &\frac 12 \bm J_{1N} \\
            \frac 12\bm J_{21}& \bm H_2  &\ldots & \frac 12 \bm J_{2N}\\
               & \ldots & \ldots & \\
            \frac 12\bm J_{N1}& \frac 12\bm  J_{N2} & \ldots &\bm H_N  \\
           \end{array}
\right) \label{eq:Hmatrix}
\end{equation}
A Gaussian distribution could be considered all bout non-problematic, at this point. However, there are a couple of good reasons to keep the system as an additive pairwise interaction model. First, although treatable, the matrix $\mathbb{H}$ could run uncomfortably large, since its size is $(L\times N)^2$, $L$ being the size of the sequences, and $N$ the number of species in the tree. Notice that  for discrete models $L$ could be particularly large, since a 21 one hot encoding is normally used to codify discrete amino acid alphabet into continuous variables, implying that $L$ is 21 times larger than the real amino acid sequence. The second reason is that dealing with a pairwise model brings to bear the statistical mechanics toolbox to compute exactly or approximate the inference problem.

\subsection{Inference problem} 
\label{sec:Inference}

In the context of ASR, tree-nodes are decomposed  in two groups, leaves and internal nodes: $\vec{X} = \{\vXh,\vXd\}$, with $\vXh$ being the internal (ancestral) nodes configurations and $\vXd$ being the leaves (data) configurations (see Fig. \ref{fig:treeXX}). The MAP estimate of ancestral sequences configurations can be computed by maximizing the posterior distribution

\begin{equation}
\label{bayes_theo_OU}
P(\vXh|\vXd) = \frac{P^0(\vec{X})}{P(\vXd)} \propto P^0(\vXh,\vXd) \propto \exp\left\lbrace -\mathcal{H}^d(\vXh)\right\rbrace 
\end{equation}
over internal sequences $\vXh$, where $P^0(\vXh,\vXd)$ from eq. (\ref{eq:ML_PairwiseDistribution}) is the equilibrium distribution for the whole system, but with the leaves evaluated at the observed values $\vXd$.

The term in the exponent, therefore, is
\begin{equation}
\label{eq:ML_Hd}
\mathcal{H}^d(\vXh) = -\sum_{1\leq i<j\leq N_h}\vec{x}^T_i \bm J_{ij}\vec{x}_j - \sum_{i=1}^{N_h} \left(\vec{x}^T_i \bm H_i \vec{x}_i + \vec{\bm h}_i \vec{x}_i\right),
\end{equation}
and  the fields and couplings $\left\lbrace \bm J_{ij}, \bm H_i\right\rbrace$ are equal to those of $\mathcal{H}^0$, and 
\begin{equation}
\label{eq:hi_def}
\bm\vec{h}_i =
\begin{cases}
\sum_{j\in c(i)}  \Lam_{ij}\Sig^{-1}_{ij} \vec{x}_j & \text{if $i$ is in contact with  leaf $\vec{x}_j$} ,\\
\vec{0} & \text{otherwise}.
\end{cases}
\end{equation}

To maximize the posterior distribution over hidden sequences  is equivalent to finding the mode of the distribution $P(\vXh|\vXd)$. The posterior  (\ref{bayes_theo_OU}) can be rewritten as: 
\begin{equation}
\label{eq:posterior_as_Gaussian}
P(\vXh|\vXd)\propto \exp\left\lbrace \vXh^{T} \mathbb{K} \vXh +\vec{S}^{T}\vXh\right\rbrace 
\end{equation}
where $\mathbb{K}$  is a block matrix, corresponding to the part of the matrix $\mathbb{H}$ (eq. (\ref{eq:Hmatrix})) that acts over the hidden nodes, with elements
\begin{equation}
\label{eq:Kmatrix}
\bm K_{ij} =
\begin{cases}
\frac{1}{2} \bm  J_{ij} & \text{if $i\neq j$} ,\\
\bm H_i & \text{if $i=j$}.
\end{cases}
\end{equation}
and   $\vec{S}$ is a concatenation of vectors $\vec{h}_i$ defined in (\ref{eq:hi_def}).

As (\ref{eq:posterior_as_Gaussian}) is a Gaussian distribution it's mode match with its mean and is given by:

\begin{equation}
\label{eq:mean_Gaussian}
\hat{\mu}=-\frac{\mathbb{K}^{-1} \vec{S}^T}{2}
\end{equation}
i.e., the mode of the posterior  distribution can be computed directly from the expression (\ref{eq:mean_Gaussian}). Unfortunately, for practical applications it implies the inversion of a matrix $\mathbb{K}$ of very high dimensions $dim(\mathbb{K})=L\times N_h $, with   $L$ the length of the sequences and $N_h$ the number of internal nodes. Gaussian message-passing algorithm (GaMP) \cite{Weiss2001,Malioutov2006,DBickson09} could be adapted to overcome this issue, which correctly solves the problem on trees, reducing  optimization to compute max-marginals

\begin{equation}
\label{max_marginals}
M_i(\vec{\mu}^{i})=\max_{\vXh}\left\lbrace P^d(\vXh): \vec{x}_i=\vec{\mu}^i\right\rbrace 
\end{equation}

Then, the set of ancestral configurations  that jointly maximizes the posterior distribution \eqref{eq:posterior_as_Gaussian} is given by $\left\lbrace \vec{\mu}^1,\vec{\mu}^2,\dots,\vec{\mu}^{N_h}\right\rbrace $. Adaptation of GaMP update rules to  compute of max-marginals yields to max-product update rules \cite{DBickson09}, details of the algorithm obtained are shown in appendix \eqref{sub:message_passing}. In what follows, we use equation \eqref{eq:mean_Gaussian} to compute the accuracy of this estimator analytically, and \eqref{max_marginals} through the GaMP algorithm, to efficiently  evaluate \eqref{eq:mean_Gaussian} without inverting the $\mathbb{K}$ matrix.

\subsection{Evaluating estimator accuracy via mean square error}

\begin{figure*}[!htb]
	\begin{center}
		\centering\includegraphics[keepaspectratio=true,width=0.6\textwidth]{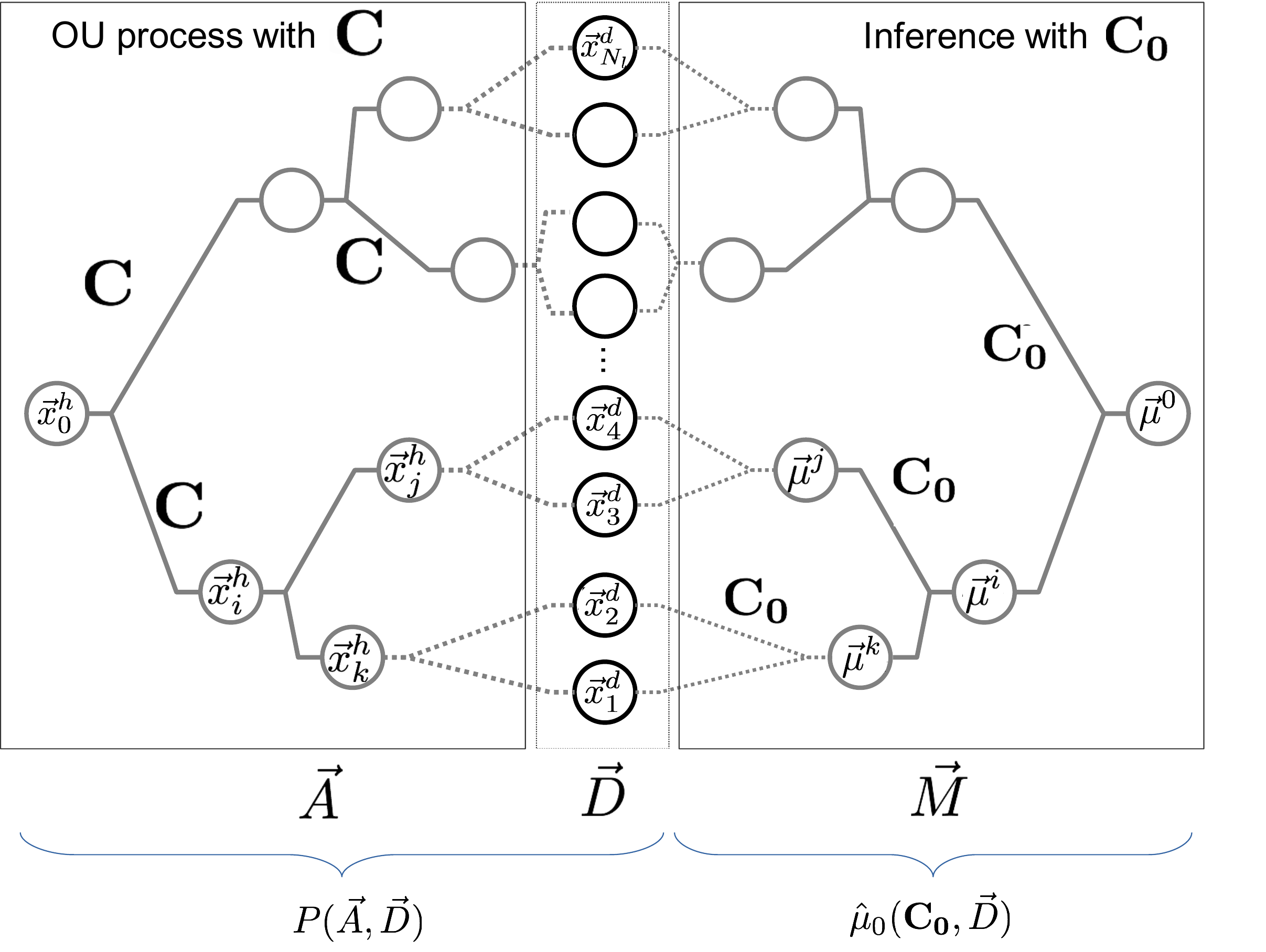}
	\end{center}
	\vspace{-1mm}
	\caption{{Schematic representation of the inference process. On the left, ancestral sequences are generated by an Ornstein-Uhlenbeck process with equilibrium covariance $\bm C$. On the right, the reconstruction of the tree is obtained by maximizing the likelihood of a similar process, albeit with a potentially different covariance $\bm C_0$. For a perfect reconstruction $\vec{M}=\vec{A}$.}} 
	\label{schemetree}
\end{figure*}

It is important to check whether the inference process defined above provides a solution that does not only maximize the posterior distribution, but that actually defines a set of inferred sequences which are similar to the actual one. We measure this similarity by the distance

	\begin{equation}
\hat{d}(\bm C,\bm C_0)=\left\langle||\vec{M} -\vXh||^2 \right\rangle_{\vec{M},\vec{A}}        =\left\langle \sum_{b}^{L\times N_h}\left( M_b-A_b\right) ^2\right\rangle_{\vec{M},\vec{A}} 
\label{square_error_general}
\end{equation}
 where  $\bm C$ encodes the information about the actual co-evolutionary process, and therefore the joint statistics of the real ancestral sequences $\vec A$ and the observed ones $\vec D$ defined by equation (\ref{eq:ML_PairwiseDistribution}) while $\bm C_0$ has a structure that is defined by the researcher before the start of the inference and is not necessarily equal to $\bm C$ as schematized in Figure \ref{schemetree}. $\vXh=\left[ \vxh_1,\dots,  \vxh_{N_{h}} \right] $  is the concatenated vector of  true ancestral sequences and  $\vec{M}=[\vec{\mu}^1\dots\vec{\mu}^{N_h}]$  are the concatenated inferred configurations at internal nodes, for a  realization  of the data $\vec{D}$.

 The estimator \eqref{eq:mean_Gaussian}  can be   rewritten  as

 $$\hat{\mu}_0(\bm C_0,\vXd)=-\frac{\mathbb{K}_0^{-1}*\mathbb{A}_0 \vXd^T}{2}$$ 
 to make explicit  it's dependence on the data $\vec{D}$ and  where $\mathbb{A} _0$ is a block matrix dependent  of the correlation matrix $\bm C_0$ and the tree topology
 \begin{equation}
 \bm {A}^{0}_{i,j} =
 \begin{cases}
 \Lam_{ij} (\bm C_0)*\Sigma^{-1}_{ij}(\bm C_0)& \text{if $i$ is an external hidden node in contact with leaves node $j$ }.\\
 \bm 0 & \text{otherwise}  
 \end{cases}
 \label{M_definition}
 \end{equation}

 We exploit different $\bm C_0$ to analyze the effect on the estimator accuracy when, for instance, a diagonal approximation of $\bm C$ is used for the reconstruction, as is the case of the independent site evolutionary models used by Pupko's algorithm \cite{pupko}. In what follows, the block matrices with subscript $0$ indicate that they are a function of the correlation matrix $ \bm C_0 $. 		 
 
In equation \eqref{square_error_general}  the sum goes over each of the $L\times N_h$ elements of the concatenated vector and the average  is done by the  joint distribution $P(\vXh,\vec{M})$ :
	
	$$\hat{d}(\bm C,\bm C_0)=\sum_b\int \left( M_b-A_b\right) ^2 P(\vXh,\vec{M}) d\vXh d\vec{M} $$

The  joint distribution $P(\vXh,\vec{M})$ can be obtained from:

\begin{equation}
P(\vXh,\vec{M})=\int d \vXd \delta [\hat{\mu}_0(\vXd)-\vec{M}] P(\vXh,\vXd) 
\label{joint_dist_general}
\end{equation}
with $P(\vXh,\vXd)$ as the full probability of all sequences: ancestral ($\vXh$) and leaves ($\vXd$) concatenated vectors. $P(\vXh,\vXd)$ was defined in \eqref{eq:ML_PairwiseDistribution}  and can be rewritten as

\begin{equation}
P^0(\vec{X})=P\left( \vXh,\vXd\right)\propto\exp\left\lbrace \vXh \mathbb{K}  \vXh^T + \vXh * \vec{S}^T-\frac{1}{2} \vXd \mathbb{G} \vXd^T\right\rbrace  
\label{full_prob_redefined}
\end{equation}
where the block matrix $\mathbb{K}$ is defined in \eqref{eq:Kmatrix} and vector $\vec{S}$ is a concatenation of vectors $\vec{h}_i$ defined in \eqref{eq:hi_def}. Finally, the elements of the block matrix $\mathbb{G} $ are given by

\begin{equation}
\bm G_{i,j} =
\begin{cases}
\bm \Sigma^{-1}_{a(i),i} & \text{ if $i=j$}.\\
\bm 0 & \text{otherwise}  
\end{cases}
\label{G_definition}
\end{equation}

Then, to evaluate equation \eqref{joint_dist_general} we use  expression \eqref{full_prob_redefined} and  the exponential representation of Dirac's delta function:

\begin{equation}
\begin{split}
P(\vXh,\vec{M})=&\int d \vXd \delta [\hat{\mu}_0(\vXd)-\vec{M}] P(\vXh,\vXd) \\
=&\int d \vXd \int d \vec{q}\exp \left\lbrace i*\vec{q}*(\hat{\mu_0}(\vXd)-\vec{M})^T\right\rbrace * P(\vXh,\vXd) \\
=& \exp\left\lbrace \vXh \mathbb{K}  \vXh^T  \right\rbrace \int d \vec{q}\exp \left\lbrace -i*\vec{q}*\vec{M}^T\right\rbrace \int d \vXd  \exp\left\lbrace -\frac{1}{2} \vXd \mathbb{G} \vXd^T+ \vXd \mathbb{A} ^T\vXh^T - \frac{i}{2} \vec{q}* \mathbb{K}_0^{-1} \mathbb{A}_0 \vXd^T \right\rbrace 
\end{split}
\label{part_1_2}
\end{equation}
such that (see details in \ref{integration}): 

\begin{equation}
P(\vXh,\vec{M})\propto\exp\left\lbrace-\frac{1}{2} \left[ -2\vXh\mathbb{K}\vXh^T+4 \vXh \mathbb{Q}_1^T\mathbb{Q}_0^{-1}\mathbb{K}_0 \vec{M}^T  +4*\vec{M} \mathbb{K}_0\mathbb{Q}_0^{-1}\mathbb{K}_0 \vec{M}^T\right] \right\rbrace 
\label{joint_dist_final_expression_2}
\end{equation}
where $\mathbb{Q}=\mathbb{A} \mathbb{G} ^{-1}\mathbb{A} ^T$ , $\mathbb{Q}_0=\mathbb{A}_0 \mathbb{G} ^{-1}\mathbb{A}_0 ^T$ and $\mathbb{Q}_1=\mathbb{A}_0 \mathbb{G} ^{-1}\mathbb{A} ^T$ .

The equation \eqref{joint_dist_final_expression_2} can be resumed as 

\begin{equation}
P(\vXh,\vec{M})\propto\exp\left\lbrace-\frac{1}{2} \vec{Z} \mathbb{V}^{-1}\vec{Z}^T\right\rbrace 
\label{joint_dist_final_expression_3}
\end{equation}
with $\vec{Z}=\left[ \vXh,\vec{M}\right] $ and 

\begin{equation} 
\mathbb{V}^{-1} =  \left(
\begin{array}{cccc}
{ -2 \mathbb{K}}&{ 2 \mathbb{Q}_1^T\mathbb{Q}_0^{-1}\mathbb{K}_0}\\
{ 2\mathbb{K}_0 \mathbb{Q}_0^{-1}\mathbb{Q}_1}& {4 \mathbb{K}_0\mathbb{Q}_0^{-1}\mathbb{K}_0 }\\\
\end{array}\label{eq:invV_2}
\right)
\end{equation}

Equation \eqref{joint_dist_final_expression_3} can be seen as a bivariate normal distribution where the variables are given by the pair of  concatenated vectors $\left[\vXh,\vec{M}\right]$. The covariance matrix for this distribution $\mathbb{V}_{ij}=\left\langle \vec{Z}_i,\vec{Z}_j\right\rangle $ allow us to compute  terms in the mean square error defined by the equation \eqref{square_error_general} if we know $\mathbb{V}_{11}=\left\langle \vXh\cdot\vXh^{T}\right\rangle $, $\mathbb{V}_{22}=\left\langle \vec{M}\cdot\vec{M}^{T}\right\rangle $ and $\mathbb{V}_{12}=\left\langle \vXh\cdot\vec{M}^T\right\rangle $. For this we invert $\mathbb{V}^{-1}$ obtaining:

\begin{equation} 
\mathbb{V} =  \left(
\begin{array}{cccc}
{ -\left[2\mathbb{K}+\mathbb{Q}\right]^{-1} }&{ \frac{1}{2}\left[2\mathbb{K}+\mathbb{Q}\right]^{-1}*\mathbb{Q}_1^T\mathbb{K}_0^{-1}}\\
{\frac{1}{2}\mathbb{K}_0^{-1}\mathbb{Q}_1\left[2\mathbb{K}+\mathbb{Q}\right]^{-1} }& {\frac{1}{4}\mathbb{K}_0^{-1}\left[ \mathbb{Q}_0-\mathbb{Q}_1\left[2\mathbb{K}+\mathbb{Q}\right]^{-1} \mathbb{Q}_1^T\right] \mathbb{K}_0^{-1}}\\\
\end{array}\label{eq:V_2}
\right)
\end{equation}

Then, we simply evaluate the expression 

\begin{equation} 
\hat{d}(\bm C,\bm C_0)=\sum_{b}^{L\times N_h} \left( \left[ \mathbb{V}_{11}\right] _{bb}-2*\left[ \mathbb{V}_{12}\right]_{bb}+\left[  \mathbb{V}_{22}\right]_{bb} \right) 
\label{error_general_2}
\end{equation}

Notice that, the mean square error from equation \eqref{error_general_2} is a function of both, the correlation matrix $\bm C$ that defines  the potential in the direct OU process and  the  matrix $\bm C_0$ which represents the approximate  correlation matrix  used for the inference. If $\bm C_0$ is diagonal, we have an independent site approximation. On the other hand, if $\bm C_0 \equiv \bm C $ we are inferring with the actual correlation for the evolutionary process.

\section{Comparison with numerical experiments}
	
In order to test our results, we extract the direct process correlation $\bm C$ randomly from a Wishart distribution:
	
$$\bm C_{L\times L}\sim W(k,\bm P)\propto \left| \bm C\right|^{(k-L-1)/2} \exp\left\lbrace -\frac{1}{2} \Tr (\bm P^{-1} \bm C)\right\rbrace $$
where parameters $k>L-1$ and $\bm P$ stand for  the number of degrees of freedom and  the scale matrix respectively.  The expected value for $\bm C$  is $E(\bm C)=k*\bm P$ and the variance of its elements is given by $$Var(C_{ij})=k*(p^2_{ij}+p_{ii}*p_{jj}) $$

If we set $\bm P=\Imat$, where $\Imat$ is the identity matrix, then $\bm C/k \rightarrow \Imat$ for $k\gg L$. This makes it possible to sample correlation matrices with different  levels of covariance by tuning the ratio $k/L$, which  allows us to understand when  it becomes  relevant  to go beyond the independent site approximation (zero off-diagonal correlation matrix).

In order to assess the relevance of neglecting the covariance of the characters in the ancestral reconstruction process, we will evaluate $\hat{d}(\bm C,\bm C_0)$  at two extreme cases: 	
\[ \hat{d}_1  = \hat{d}(\bm C,\bm C)  \qquad  \hat{d}_0=\hat{d}(\bm C,\bm C_0= diag (\bm C )) \]
where in the second case the inference process is carried out with a factorized assumption on the distribution of the characters, and therefore with a diagonal $\bm C_0$. For a fixed $\bm C$, both $\hat{d}_1$ and $\hat{d}_0$ are monotonically increasing functions of the speed of the evolutionary process $\gamma$ (not shown in figures), starting from $\hat d = 0$ when $\gamma=0$ up to the average distance between two uncorrelated equilibrium configurations when $\gamma = \infty$, with $\hat{d}_0 > \hat{d}_1$ at every $\gamma$, as expected.

\begin{figure*}[!htb]
	\begin{center}
		\centering\includegraphics[keepaspectratio=true,width=0.6\textwidth]{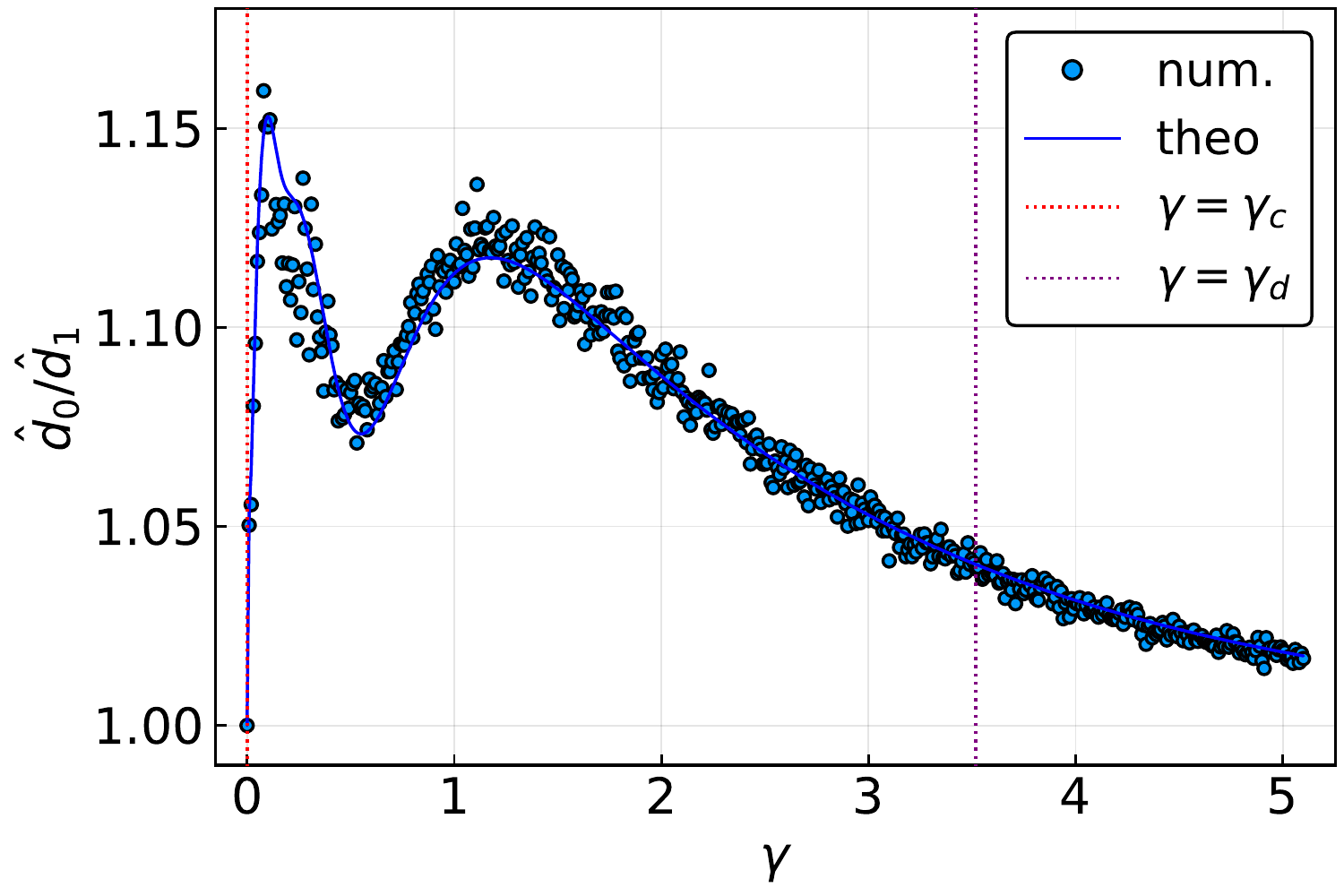}
	\end{center}
	\vspace{-1mm}
		\caption{{Ratio between $\hat{d}_0$ and $\hat{d}_1$ as function of the parameter $\gamma$ for a binary tree  and homogeneous topology with $\Delta t=1.0$ and height $h=7$. The OU potential $\bm C$ is sampled from a Wishart distribution with  $\bm P=\Imat$ and $k=L=10$.  Solid  line corresponds to the evaluation of the equation \eqref{error_general_2}, whereas points are the result of averaging 100  numerical simulations of the evolution and  inference (using GaMP) for each vale of $\gamma$. The verticals dashed lines represent the borders of the interval for time-scale parameter where the inference results non-trivial.  }} 
	\label{d0_d1_full_tree_8_L10_k_eq_p_Wishart}
\end{figure*}

In Figure \ref{d0_d1_full_tree_8_L10_k_eq_p_Wishart} is shown the ratio between $\hat{d}_0=\hat{d}(\bm C,\bm C_0= diag (\bm C ))$  and $\hat{d}_1=\hat{d}(\bm C,\bm C_0=\bm C)$  as function of the time-scale parameter $\gamma$ for a binary tree with height $h=7$ and for a direct potential $\bm C$  sampled from a Wishart distributions where $k=L=10$. The plot was also reproduced numerically using GaMP from 100 simulations of the evolution-inference process for each $\gamma$ value. 

Not surprisingly, extreme cases for the typical time-scale $\gamma = 0$ and $\gamma = \infty$ produce no difference ($\hat{d}_1 = \hat{d}_0$) in the accuracy of the estimator using either  $\bm C$ or $\bm C_0$. One of the cases $\gamma \to 0$ means that the process is too slow to produce any changes in the sequences along the tree, and all observed leaves and all hidden nodes have the same value and can be equally (trivially and perfectly) reconstructed independently of the covariance assumed. The other case, $\gamma \to \infty$ corresponds to an extremely fast evolutionary process, such that all nodes are equilibrium samples, and there is no information whatsoever in the observed data. In this limit  both methods are equally bad.

 We can find effective values for the parameter $\gamma$ where the two regimes described above  start to be  noticeable. We know that  the correlation between two sequences linked by an OU process is given by:

$$\Lam^{\Delta t} \bm C = \sum_{a}^{L} \rho_a^{-1} e^{-\rho_a \Delta t* \gamma}  \left| s_a\right\rangle \left\langle s_a\right| =\sum_{a}^{L} \rho_a^{-1} e^{- \Delta t /\tau_a}  \left| s_a\right\rangle \left\langle s_a\right|$$
with $\left\lbrace \rho_a, \left\langle s_a\right|\right\rbrace $ eigenstates of $\bm C^{-1}$  and $\tau_a=(\gamma*\rho_a)^{-1}$, then we can consider that two configurations are \textit{uncorrelated}  when $\Delta t \gg \max(\tau_a)=\frac{1}{\gamma*\rho_{min}}$ or $\gamma\gg \frac{1}{\Delta t \rho_{min}}$. Inversely an \textit{strongly correlated} regime occur when $\Delta t \ll \min(\tau_a)=\frac{1}{\gamma*\rho_{max}}$ or $\gamma\ll\frac{1}{\Delta t \rho_{max}}$. These criteria could be generalized for a tree if we set $\Delta t=\Delta t_{av}$ where $\Delta t_{av}$ is the average time between connected nodes in the tree. Then we have $ \gamma\gg \frac{1}{\Delta t_{av}\rho_{min}}=\gamma_d$ for \textit{uncorrelated} regime and $ \gamma\ll\frac{1}{\Delta t_{av}\rho_{max}}=\gamma_c$ for the \textit{strongly correlated} regime.

The interval where  phylogenetic-based inference methods are relevant is  \[  \frac{1}{\Delta t_{av}\rho_{max}}\equiv  \gamma_c   \ll \:\:  \gamma  \:\: \ll \gamma_d \equiv  \frac{1}{\Delta t_{av}\rho_{min}} \]

In this regime, the data is neither too correlated (\textit{strongly correlated} regime) around the tree as to make inference unnecessary or too uncorrelated, making the phylogeny irrelevant  (\textit{uncorrelated} regime).  In Figure \ref{d0_d1_full_tree_8_L10_k_eq_p_Wishart} this interval is signaled with vertical lines, and it coincides with the region where the full correlated nature of the process more significantly outperforms the factorized inference $\frac{\hat{d}_0}{\hat{d}_1}>1$.

\begin{figure*}[!hb]
	\begin{center}
		\centering\includegraphics[keepaspectratio=true,width=0.6\textwidth]{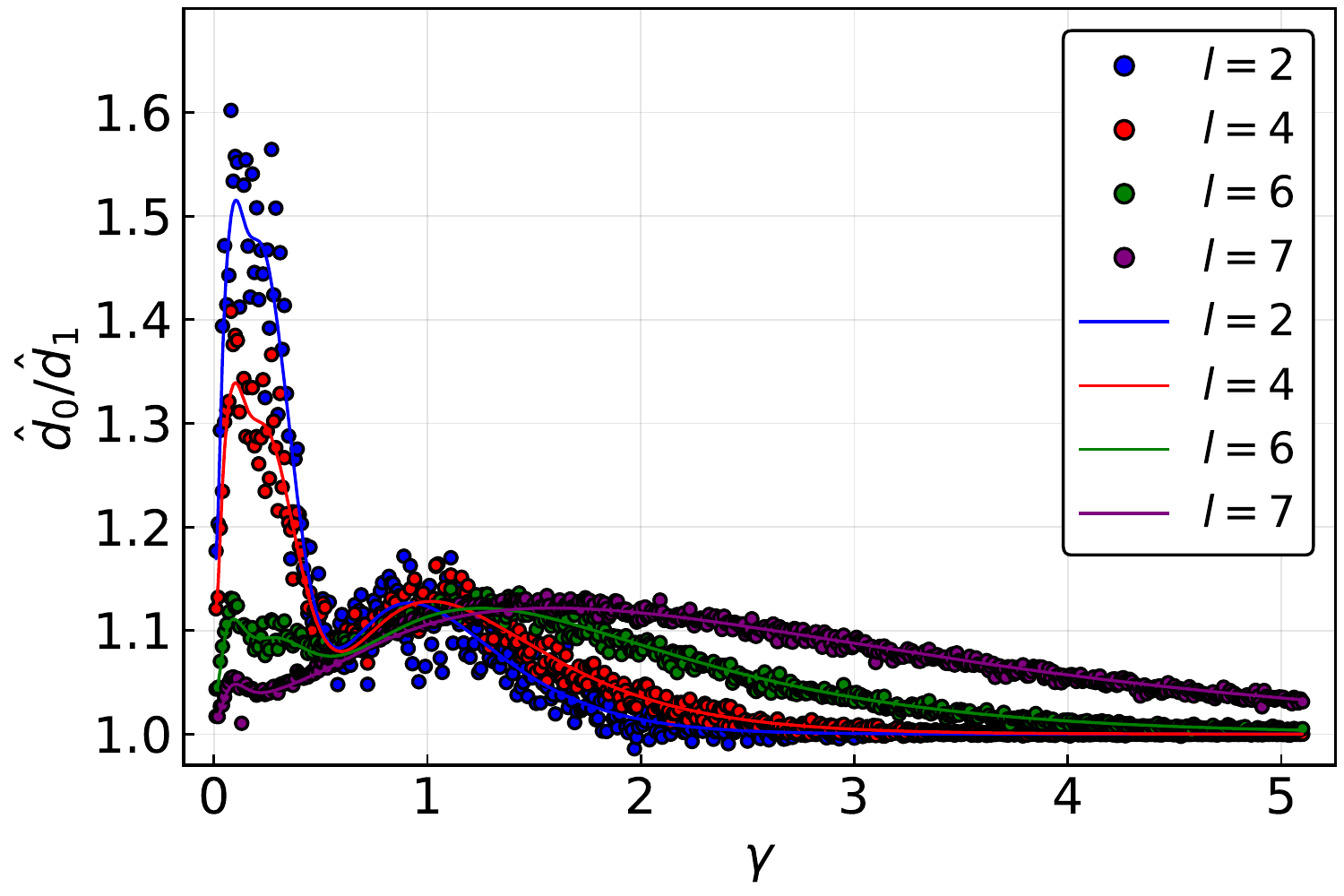}
	\end{center}
	\vspace{-1mm}
	\caption{{ Ratio between $\hat{d}^l_0$ and $\hat{d}^l_1$  as function of the parameter $\gamma$ for different levels $l$, for a tree with binary and homogeneous topology with $\Delta t=1.0$ and height $h=7$. The OU potential $\bm C$ is sampled from a Wishart distribution with  $\bm P=\Imat$ and $k=L=10$. Solid  line correspond to the evaluation of the equation \eqref{error_general_2} but using in place of $\mathbb{V}$ a matrix formed only by blocks associated with the level $l$.   Points are the result of averaging 100  numerical simulations of the evolution and  inference (using GaMP) for each vale of $\gamma$.  }}
\label{d0_d1_levels_tree_8_L10_k_eq_p_Wishart}
\end{figure*}

Notice that the ratio has two maximums that are originated by different mechanisms. The first one reflects the value of $\gamma$ at which the system instantly recognizes de difference between the two matrices in the expressions pre-multiplied by the factor $\gamma$ as $\Lam=\exp{-\gamma \bm C^{-1}}$ and $\Lam_0=\exp{-\gamma \bm C_0^{-1}}$. This effect has different proportions at each level in the tree as is shown in Figure \ref{d0_d1_levels_tree_8_L10_k_eq_p_Wishart} where we plot the ratio between the inference errors but for  different levels of the tree  $\hat{d}^l_0$ y $\hat{d}^l_1$. These errors are computed as in equation \eqref{error_general_2}, but using in place of $\mathbb{V}$ a matrix formed only by the blocks associated with the corresponding level $l$.  For internal nodes, the effective value of $\gamma$ is higher as soon as the node is more distant from the leaves.

At larger values of $\gamma$, the fraction $\frac{\hat{d}_0}{\hat{d}_1}$ starts to grow with $\gamma$ because the differences between  $\Lam$ and $\Lam_0$ becomes more important. This growth saturates at the second peak of the plot, from this point  the system  moves toward the \textit{uncorrelated} regime and $\Lam\approx\Lam_0\rightarrow0$. Again, this exponential decay from the maximum emerges first for more internal levels of the tree.

\newpage
\section{Sequences with discrete characters}
\label{Monte_Carlo_dynamics}

Unfortunately, when dealing with sequences with discrete characters, we don't have a global co-evolutionary propagator for the statistical model of equation  \eqref{Potts}. Then, it's impossible to derive and explicit form of the  joint probability of ancestral configurations given the contemporary sequences of equation \eqref{bayes_theo_OU}.  However, instead of the common factorization of the propagator, we may keep the global nature and tractability of the problem transforming the Potts model over discrete variables \eqref{Potts} into a Gaussian distribution, and to assume an  OU dynamics as  an approach to the evolutionary process.   In this way, we may exploit  the same inference scheme described for continuous traits in \ref{sec:Inference}. In what follows, we expand on this idea. 

We know that it is possible to represent a Potts model (\ref{Potts}) unambiguously in the space of the frequencies from the correlation matrix  $\left\lbrace \bm C\right\rbrace $. In this space we can also define a multivariate Gaussian model that describes the same statistics of the discrete model but for continuous variables \cite{Baldassi}. In practice, from a set of $M$ sequences sampled by the pairwise Potts model,  we first transform each sequence of length $L$ and alphabet $\left\lbrace 1,...q\right\rbrace $ toward an array of length $q*L$ where each element take values on  binary alphabet $\left\lbrace 0,1\right\rbrace $. Then, each original sequence site is mapped to $q$ binary variables, taking  value $1$ if the position corresponds with the original state of the residue and $0$ for the rest of the positions. The result is a new  alignment with ones and zeros keeping the same structure from the original alignment but now with  dimensions $M\times(q*L)$. These binary variables could be approximated by real valued variables, which allow the computation of the covariance matrix $\bm C$ needed to parametrize the multivariate Gaussian distribution by 
\begin{equation}
P^0(\vec{x}) \propto \exp-\frac{1}{2}\left\{ \vec{x}^T\bm{C}^{-1}\vec{x} \right\}
\label{gauss_multiv}
\end{equation}

The Gaussian distribution obtained can  be easily propagated using an  Ornstein-Uhlenbeck process.  That is, we can  build from the potential $\bm C$,  the propagator $P^0(\vec{x}_j|\vec{x}_i;\Delta t_{ij})$ given by the  equation \eqref{eq:propagator_OU},  and  use it  as the propagator of the Potts model. This propagated statistics is expected to be  comparable to the one obtained from simulating the original Potts model with a Monte Carlo (MCMC) procedure.

Then, we could ask  whether it is possible to reconstruct the discrete configurations of the internal nodes of the tree if the direct evolutionary process is carried out by a Monte Carlo  dynamics for a  Potts' model \eqref{Potts}.  Pott's parameters $\bm \lambda$, for sequences with length $L$ and alphabet size  $q$,  are chosen  randomly from a Gaussian distribution with mean $\mu_{\bm \lambda}>0$ and   standard deviation $\sigma_{\bm \lambda}\ll\mu_{\bm \lambda}$. The connectivity $c$ between sites of the sequences could take different values in order to simulate  regimes with different  covariance relevance.

Given a tree structure, the direct evolutionary process starts sampling the root configuration from \eqref{Potts} via MCMC.  Then, each bifurcation event in the tree starts from the  configuration at the ancient node  and on average  $\Delta t*L*\delta$  MCMC changes  are proposed with Metropolis acceptance rate, leading to a new  configuration at the child node. Here $\Delta t$ is the tree branch length between two nodes and $\delta$ is the time-scale for the MCMC simulation from which we regulate  the phylogeny.  From sampling tree-nodes via MCMC  we obtain at leaves the discrete  data configurations $\left\lbrace \underline{X}^m\right\rbrace $  and as internal nodes  sequences which we would like to predict $\left\lbrace \underline{A}^m\right\rbrace $.

As we already discussed in the introduction, most ASR algorithms of the literature assume that sequences follow a single-site evolutionary model. Within this class, we will use an efficient dynamic programming algorithm  developed by Pupko et al.\cite{pupko} (included in FastML program) as a benchmark to compare our results. For this, we use the independent  site model of  evolution, introduced by Felsenstein \cite{Felsenstein}, with constant mutation rate $\mu$ given by

\begin{equation}
P_i(A_1|A_2,\Delta t)=e^{-\mu \Delta t}\delta_{A_1,A_2}+\left( 1-e^{-\mu \Delta t}\right) P_i(A_1)
\label{single_site_propagator}
\end{equation}

This model describes, in a time interval $\Delta t$, no mutations with probability $e^{-\mu \Delta t}$ and one or more mutations with probability $1-e^{-\mu \Delta t}$. In this last case, a specific  character states is selected according $P_i(A_1)$. Details of our implementation of the algorithm appear in appendix \eqref{ap:parameter_FastML}. In the next section, we are going to show results for the reconstruction process  carried out with both, the  FastML algorithm and the continuous OU approximation discussed above.

Notice that the OU propagator  needs both $\bm C$ and $\gamma$. $\bm C$  is obtained from an   i.i.d sample of the Potts model,  which is  transformed  to its continuous version as described  above. On the other hand, the time-scale parameter $\gamma$ is inferred  by maximizing the likelihood of the data points at leaves of the tree as is described in appendix \eqref{Inferring gamma}.

\subsection{Numerical Results}

We generate data via  MCMC simulations, as was described in the previous section. For this, we consider a tree of binary and homogeneous topology  with  $h=9$ bifurcations events. For a system with   $L=10$ and  $q=2$,  a Potts model of connectivity  $c=3$  is designed with ferromagnetic couplings Gaussian distributed with mean $\mu_J=0.8$ and standard deviation $\sigma_J=0.2$. For simulations, we  use different  time-scale $\delta$ in the range $[0.2,11.2]$, which allows us to explore different regimes of correlation between nodes configurations. To avoid statistical noise, we repeat the sampling procedure 100 times for each value of $\delta$. Then we applied two different reconstructions strategies:  continuous OU approximation  with either the full covariance matrix $\bm C$ (MP1) or  with a matrix resulting of neglect sites covariance  $diag(\bm C)$ (MP0)  and Pupko's ML strategy with Felsenstein evolutionary model of equation \eqref{single_site_propagator} (FastML).

\begin{figure*}[!htb]
	\begin{center}
		\centering\includegraphics[keepaspectratio=true,width=0.7\textwidth]{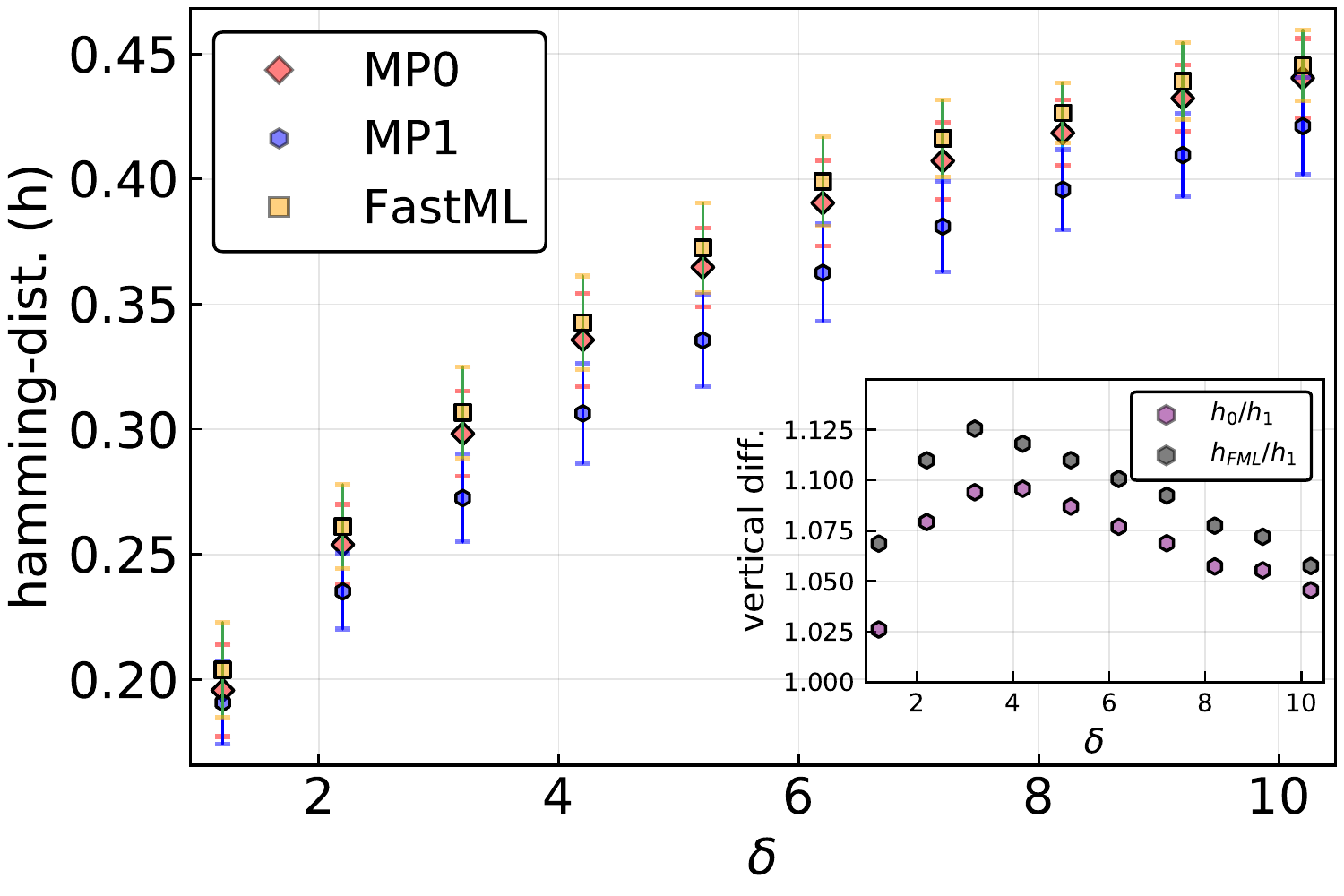}
	\end{center}
	\vspace{-1mm}
	
	\caption{{ Mean hamming distance between all  true sequences and those obtained after the inference method. Each point correspond to 100 sampling-inference process averaged.  The length of the sequence vectors was $L =10$  and the alphabet size $q=2$, the average site connectivity is $c=3$, the number of ancestral sequences $M = 511$ as consequence of a binary non-homogeneous tree with $K = 9$ bifurcations processes. Label MP0 correspond to the case when inference is carried out using continuous approximation  neglecting covariance between sites, MP1   to the case when the inference is done using the full correlation matrix $\bm C$ and  FastML is  the reference algorithm for  maximum-likelihood reconstruction developed by Pupko \cite {pupko}. The inserted plot indicates the ratio between hamming distances from MP0 or  FastML with hamming distances from MP1 approach. }} 
	\label{hd_vs_delta_three_methods}
	
\end{figure*}

In Figure \ref{hd_vs_delta_three_methods} we present  results of the different strategies.  We use the hamming distance between real and inferred ancestral sequences to evaluate the performance of the predictions. As can be seen, the performance between the different no-correlated approximations (MP0 and FastML) almost coincide, while MP1 is consistently better than both approximations. The ratios between the hamming distances obtained from  methods MP0 or FastML with hamming distances from MP1 are shown in the inset plot. This illustrates that the major gain in the prediction came from using  MP1 instead of FastML and  at intermediate values of $\delta$ where node configurations have enough divergence but are still correlated.

\newpage
\section{Conclusions}
\label{sec:discussion}

In this work, we have studied the reconstruction of ancestral sequences described by global co-evolutionary models and study  the impact of the intra-species correlations in the performance of the inference process. We studied two types of sequences. Sequences of continuous variables that evolved according to an Ornstein-Uhlenbeck dynamics on a finite tree, and sequences of discrete variables defined by a Potts model and sampled on  trees via Monte Carlo simulations. Exploiting the Gaussian character of the Ornstein-Uhlenbeck process, we were able to design a fast algorithm that provides a MAP estimate of the ancestral sequences and takes into account the covariance between variables, i.e. our algorithm goes beyond the standard independent site approximation. We were also able to analytically quantify the precision of our reconstruction and showed that this analytical computation describes correctly the results from the algorithm in artificial data sets. Then, exploiting a known projection of the Potts model on a multivariate Gaussian distribution, we tested our algorithm  in artificial discrete data sets. Also, for this kind of models, our results support the idea that there is a wide range of parameters for which the intra-species correlation significantly affect the inference process. We finally show that in this regime, our algorithm outperforms a traditional reconstruction method based on the independent site approximation.

A different, but connected issue, is the ability to distinguish, from a sequence alignment, which correlations are originated by phylogeny and which ones are originated by epistasis. In the context of Inverse Statistical Physics of biological sequences,  global co-evolutionary models are used to describe sequence variability in ensembles of homologous sequences. This allows to unveil statistical constraints acting on this variability and relate them to biological features. Potts models in Direct Coupling Analysis (DCA) have  found widespread application in protein-structure prediction from sequences \cite{morcos2011direct,cocco_inverse_2018}.  However, one of the basic assumptions of this global statistical modeling is that sequences form an, at least, approximately independent sample of an unknown probability distribution, which is to be learned from data. In the case of protein families, this assumption is obviously violated by phylogenetic relations between protein sequences. In \cite{horta2021phylogenetic} it was shown that phylogenetic correlations between sequences lead to a changed residue-residue correlation structure, represented by a fat tail in the eigenvalue spectrum of the data covariance matrix. Furthermore, the phylogeny has a global influence on all the parameters of the model and it could impact in those applications that use DCA as a sequence model. 
	
Notice that, the  equilibrium distribution reached  by the OU process is  the  Gaussian version of the DCA model, where the covariance matrix $\bm C$ is a quadratic potential that represents selection forces.  Then, in the light of  results of \cite{horta2021phylogenetic}, we must consider  that when we model biological sequences, the phylogenetic correlations interfere in the covariance signal empirically estimated from data and this not properly represent the phenotipic constrains resulting from natural selection. It is therefore necessary to disentangle covariances in the data corresponding to their multivariate Gaussian equilibrium distribution from those resulting from the historical correlations. To solve this problem \cite{erh2021} developed  a methodology which leads to a clear gain in accuracy in the inferred equilibrium distribution. However, in our  ancestral reconstruction method, we assume that this covariance matrix between residues is known. A future step is to couple both inference methods.
\\
\textbf{Acknowledgments:} 
We acknowledge interesting discussions with Martin Weigt and Pierre Barrat-Charlaix. This project has received funding from the European Union’s Horizon 2020 research and innovation program MSCA-RISE-2016 under Grant Agreement No. 734439 INFERNET. This research has also been funded by the Office for the management of funds and projects of the Ministry of Science, Technology and Environment of the Republic of Cuba within the project PN223LH010-015.

\clearpage
\bibliography{ASR}
\bibliographystyle{unsrt}

\newpage
\appendix

\section{Description of technical details} 
\label{sec:supplementary_material}
\subsection{Ornstein-Uhlenbeck dynamics}
\label{app:OU}
Let us consider a system characterized by $L$ continuous degrees of freedom and whose state is fully described by an $L$-dimensional vector $\vec{x}\in \mathbb R^L$. We suppose that this system evolves under  the potential $V(\vec{x}) =\frac{1}{2} \left\lbrace \vec{x}^T \bm J \vec{x}\right\rbrace $ being  $\bm J$  a symmetric and positive definite coupling matrix,  and according to the   Langevin equation  

\begin{equation}
\gamma^{-1} \frac{\ddroit \vec{x} }{\ddroit t} = - \bm J \vec{x}  + \vec{\xi}(t)
\label{eq:langevin}
\end{equation}
which  represents  a multivariate Ornstein-Uhlenbeck   process, where $ \vec{\xi}(t)$ is an stochastic term, and $\gamma^{-1}$   is the characteristic timescale governing the dynamics. Modeling the stochastic term $\vec\xi(t)$ as a uncorrelated white noise, we obtain  the Ito stochastic differential equation for a multivariate OU process \cite{gardiner,singh2017multiOU}:
\begin{equation}
\ddroit x_i = -\sum_{j=1}^{L}\left(  \gamma J_{ij}x_j dt + \left( \sqrt{2 D}\right) _{ij} dW_j\right) 
\end{equation}

where:
\begin{itemize}
	\item $dW_j=\xi_j(t) dt$ represent a stochastic Wiener process. 
	\item $D_{ij}=\gamma\delta_{ij}$ is a  matrix of diffusions coefficients.  
\end{itemize}

It can be shown that the corresponding Fokker-Planck equation is

\begin{equation}
\partial_t P_{1|1}=\hat{L} P_{1|1} 
\end{equation}
where  $P_{1|1}\equiv P( \vec{x}, t|\vec{x}_0, t_0)$ is the probability density of displacement from $ \vec{x}_0$ at time $t_0$ to $ \vec{x} $ at time $t$, and $\hat{L} $ is the  Fokker-Planck operator given by:

\begin{equation}
\hat{L}( \vec{x})=-\gamma \left( \sum_{i,j=1}^{L}\frac{\partial}{\partial x_i}J_{ij} x_j+\sum_{i=1}^{L}\frac{\partial^2}{\partial x^2_i}\right) 
\end{equation}
The stationary solution for the Fokker-Planck equation $\hat{L} P^0( \vec{x})=0$ is:
\begin{equation}
P^0( \vec{x}) = \frac{1}{\sqrt{(2\pi)^L \vert \bm{C}\vert}}\exp\left\{ -\frac{1}{2} \vec{x}^T\bm{C}^{-1} \vec{x} \right\}
\label{eq:stat}
\end{equation}
which is a zero mean Gaussian distribution with $\bm C=\left\langle  \vec{x}  \vec{x}^T\right\rangle $ as the covariance matrix. If from the beginning we add to the potential  $V(\vec{x})$ a  linear term $\vec{h}\cdot\vec{x}$ with $\vec{h}$ a local field, we would obtain a Gaussian distribution with mean value shift from zero and covariance matrix $\bm C=\left\langle \vec{x} \vec{x}^T\right\rangle- \left\langle \vec{x} \right\rangle \left\langle \vec{x} \right\rangle$.

The solution of the Fokker-Planck equation is a multivariate normal distribution

\begin{equation}
P(\vec{x}  | \vec{x}_0 , \Delta t) =\left[ (2\pi)^L \det \bm\Sigma\right] ^{-1/2}\exp\left\lbrace-\frac{1}{2}(\vec{x}-\vec{\mu})^T \bm\Sigma^{-1}(\vec{x}-\vec{\mu}) \right\rbrace 
\end{equation}

where 
$$ \Lam = e^{-\gamma \bm J }, \qquad\vec{\mu} = \Lam^{\Delta t} \vec{x}_0, \qquad \bm{\Sigma} = \bm{C} - \Lam\bm{C}\Lam, \qquad \Delta t=t-t_0$$ 
Matrices $\bm J$, $\bm C$ and $\bm D$ are not independent as they are related by  Liapunov stationary condition :
$$\bm J \bm C+(\bm C \bm J)^T=2 \bm D$$

from  which we get

\begin{equation}
\bm J=\bm C^{-1}
\end{equation}

Then we can conclude that the system  evolves under a quadratic potential  $V(\vec{x})=\frac{1}{2}\vec{x}^T\bm{C}^{-1}\vec{x} = \frac{1}{2}\sum_{i,j} J_{ij}x_i x_j$ and the evolutionary process leads to stationary distribution  \ref{eq:stat} when $\gamma\Delta t>>1$.

Another important property  of the Ornstein-Uhlenbeck process is  that the times correlation function obeys the linear regression theorem

\begin{equation}
\bm G(t_1-t_2)\equiv \left\langle \vec{x}_1(t_1)\vec{x}_2(t_2)\right\rangle= \Lam^{\Delta t} \bm{C}
\end{equation}

describing the covariance of configurations $\vec{x}_1$ and $\vec{x}_2$ separated in time by $\Delta t=t_1-t_2$.

\subsection{ Message passing for continuous variables } 
\label{sub:message_passing}
Posterior probability distribution \eqref{bayes_theo_OU} can be written from the following pairwise factorization 

\begin{equation}
\label{pairwise factorization }
P^d(\vXh|\vec{D})\propto \prod_{i=1}^{N_h} \phi_i(\vec{x}_i) \prod_{1\leq i<j\leq N_h} \psi_{ij}(\vec{x}_i,\vec{x}_j)
\end{equation}
where $\psi_{ij} = \exp \left( \vec{x}_i \bm J_{ij}\vec{x}_j \right)  $ and $\phi_i(\vec{x}_i)=\exp \left( \vec{x}_i \bm H_i\vec{x}_i + h_i \vec{x}_i\right) $ are edge potentials and self potential which define the tree-graph. 

Message from $i$ to $j$ over their shared edge on the tree is given by, 
\begin{equation}
\label{message_cont}
m_{ij}(\vec{x}_j) \propto\int \ddroit \vec{x}_i  \psi_{ij}(\vec{x}_i,\vec{x}_j)\phi_i(\vec{x}_i)\prod_{k\in\partial i\setminus j}m_{ki}(\vec{x}_i)
\end{equation}

and marginals 

\begin{equation}
\label{beliefs}
M_i(\vec{x}_i) \propto\phi_i(\vec{x}_i) \prod_{k\in\partial i}m_{ki}(\vec{x}_i)
\end{equation}

\subsubsection{Gaussian message passing algorithm \cite{DBickson09}} 
From previous expressions, we note that both binary and unary potentials are Gaussian :

\begin{equation}
\label{potentials}
\phi_i(\vec{x}_i)\propto N(\vec{\mu}_{i}=\frac{-\vec{h}_i}{2*\bm H_i},\bm \sigma_i=-\frac{1}{2*\bm H_i})
\end{equation}
Since Gaussian densities' product over a common variable is, up to a constant factor, also a Gaussian density, we can write messages in the following way
\begin{equation}
m_{ij}(\vec{x}_j) \propto N(\vec{\mu}_{ij},\bm \sigma^{-1}_{ij} )
\end{equation}

and the product 

\begin{equation}
\phi_i(\vec{x}_i) \prod_{k\in\partial i\setminus j}m_{ki}(\vec{x}_i) \propto N(\vec{\mu}_{i\setminus j},\bm \sigma^{-1}_{i\setminus j} )
\end{equation}

is also Gaussian with

\begin{equation}
\begin{split}
\bm \sigma^{-1}_{i\setminus j}&=\bm \sigma^{-1}_{i}+\sum_{k\in\partial i\setminus j} \bm \sigma^{-1}_{ki}\\
\vec{\mu}_{i\setminus j}&=\bm \sigma_{i\setminus j}  \left( \bm \sigma^{-1}_{i} \vec{\mu}_i+\sum_{k\in\partial i\setminus j} \bm \sigma^{-1}_{ki}\vec{\mu}_{ki}\right) 
\end{split}
\end{equation}

Plugging this into the message defining equation \ref{message_cont}, we obtain

\begin{equation}
\label{gaussian_mij}
\begin{split}
m_{ij}(\vec{x}_j) &\propto\int \ddroit \vec{x}_i  \psi_{ij}(\vec{x}_i,\vec{x}_j)\phi_i(\vec{x}_i)\prod_{k\in\partial i\setminus j}m_{ki}(\vec{x}_i)\\
&\propto \int\ddroit \vec{x}_i \exp(\vec{x}_i \bm J_{ij} \vec{x}_j ) \exp(-\frac{1}{2}\vec{x}_i \bm \sigma^{-1}_{i\setminus j} \vec{x}_i + \sigma^{-1}_{i\setminus j} \vec{\mu}_{i\setminus j} \vec{x}_i )\\
&=\int\ddroit \vec{x}_i  \exp(-\frac{1}{2} \vec{x}_i \bm \sigma^{-1}_{i\setminus j} \vec{x}_i + (\bm \sigma^{-1}_{i\setminus j} \vec{\mu}_{i\setminus j} +\bm  J_{ij}\vec{x}_j ) \vec{x}_i )\\
&\propto \exp ( (\bm \sigma^{-1}_{i\setminus j}\vec{\mu}_{i\setminus j}+\bm J_{ij}\vec{x}_j)^2  / (2*\bm\sigma^{-1}_{i\setminus j}))\\
&\propto N(\vec{\mu}_{ij}=\bm \sigma_{ij} \bm J_{ij} \vec{\mu}_{i\setminus j},\bm \sigma_{ij}=-\bm \sigma_{i\setminus j}^{-1}/(\bm J_{ij})^2)
\end{split}
\end{equation}
leading to the update rules for Gaussian parameters:

\begin{equation}
\label{update rules_mij}
\begin{split}
\bm \sigma^{-1}_{ij} =& -\bm J^2_{ij}/(\bm \sigma^{-1}_i+\sum_{k\in\partial i	\setminus j} \bm\sigma^{-1}_{ki})\\
\vec{\mu}_{ij} =&- (\vec{\mu}_i \bm\sigma^{-1}_{i} +\sum_{k\in\partial i\setminus j}\bm\sigma^{-1}_{ki}\vec{\mu}_{ki})/\bm J_{ij} 
\end{split}
\end{equation}

substituting \ref{potentials} and \ref{gaussian_mij} in \ref{beliefs}  we obtain the marginals as a Gaussian density $M_i(\vec{x}_i)\propto N(\vec{\eta}_i,\bm\kappa_i)$  with

\begin{equation}
\label{marginals_param}
\begin{split}
\vec{\eta}_i =& (\bm\sigma^{-1}_{i} \vec{\mu}_{i}+\sum_{k\in\partial i}\bm \sigma^{-1}_{ki}\vec{\mu}_{ki}) /(\bm\sigma^{-1}_{i} +\sum_{k\in\partial i}\bm\sigma^{-1}_{ki})\\
\bm \kappa_i^{-1} =& \bm\sigma^{-1}_i + \sum_{k\in\partial i}\bm\sigma^{-1}_{ki}
\end{split}
\end{equation}

Update rules equation in \ref{marginals_param} match with  ones associated to the max-product algorithm, as we will show below. This means that our solution in \eqref{max_marginals} can be obtained as 
$$\vec{x}^*_i=\vec{\eta}_i$$

\subsubsection{Max-Product rule} 

A continuous version of max-product algorithm could be obtained, replacing the integral-product rule by 
\begin{equation}
\label{max-product_mij }
m_{ij}(\vec{x}_j) \propto \arg \max_{\vec{x}_i}\psi_{ij}(\vec{x}_i,\vec{x}_j)\phi_i(\vec{x}_i)\prod_{k\in\partial i\setminus j}m_{ki}(\vec{x}_i)
\end{equation}

similar to \ref{gaussian_mij} we get 

\begin{equation}
\label{max_sum_messages}
\begin{split}
m_{ij}(\vec{x}_j) &\propto \arg \max_{\vec{x}_i}  \exp(\vec{x}_i \bm J_{ij} \vec{x}_j ) \exp(-\frac{1}{2}\vec{x}_i \bm\sigma^{-1}_{i\setminus j} \vec{x}_i + \bm\sigma^{-1}_{i\setminus j} \vec{\mu}_{i\setminus j} \vec{x}_i )\\
&=\arg \max_{\vec{x}_i}   \exp(-\frac{1}{2} \vec{x}_i \bm\sigma^{-1}_{i\setminus j} \vec{x}_i + (\bm\sigma^{-1}_{i\setminus j} \vec{\mu}_{i\setminus j} + \bm J_{ij}\vec{x}_j ) \vec{x}_i )
\end{split}
\end{equation}
deriving and equating to zero the exponential term, we find 

\begin{equation}
\vec{x}^{max}_i=\frac{\bm\sigma^{-1}_{i\setminus j}\vec{\mu}_{i\setminus j}+\bm J_{ij}\vec{x}_j}{\bm\sigma^{-1}_{i\setminus j}}
\end{equation}

substituting $\vec{x}^{max}_i$ back in \ref{max_sum_messages} we get, 

\begin{equation}
\begin{split}
m_{ij}(\vec{x}_j) &\propto \exp (     (\bm\sigma^{-1}_{i\setminus j}\vec{\mu}_{i\setminus j}+\bm J_{ij}\vec{x}_j)^2  / (2*\bm\sigma^{-1}_{i\setminus j}))\\
&\propto N(\vec{\mu}_{ij}=\bm\sigma_{ij} \bm J_{ij} \vec{\mu}_{i\setminus j},\bm\sigma_{ij}=-\bm\sigma_{i\setminus j}^{-1}/(\bm J_{ij})^2)
\end{split}
\end{equation}
which is identical to the messages derived for the sum-product case, then as intuitively we could guess the rules obtained to find the marginals (\ref{update rules_mij}  and\ref{marginals_param}) leads to max marginals for the Gaussian version of Message Passing and then  as marginals are Gaussian the maximum value correspond to the mean.

\newpage

\subsection{Integration of equation \ref{part_1_2}}
\label{integration}

We must first evaluate:
\begin{equation}
\begin{split}
I_3=&   \int d \vXd  \exp\left\lbrace -\frac{1}{2} \vXd \mathbb{G} \vXd^T+ \vXd \mathbb{A} ^T\vXh^T - \frac{i}{2} \vec{q}* \mathbb{K}_0^{-1} \mathbb{A}_0 \vXd^T \right\rbrace  \\
=&cte*\exp \left\lbrace \frac{1}{2} \left(\vXh\mathbb{A}-\frac{i*\vec{q} \mathbb{K}_0^{-1} \mathbb{A}_0}{2}\right)  \mathbb{G} ^{-1}\left( \vXh\mathbb{A}-\frac{i*\vec{q}\mathbb{K}_0^{-1} \mathbb{A}_0}{2}\right)^T\right\rbrace 
\end{split}
\end{equation}

defining $\mathbb{Q}=\mathbb{A} \mathbb{G} ^{-1}\mathbb{A} ^T$ , $\mathbb{Q}_0=\mathbb{A}_0 \mathbb{G} ^{-1}\mathbb{A}_0 ^T$ and $\mathbb{Q}_1=\mathbb{A}_0 \mathbb{G} ^{-1}\mathbb{A} ^T$  we obtain:

\begin{equation}
I_3=cte*\exp \left\lbrace \frac{1}{2} \left[ \vXh\mathbb{Q}\vXh^T-i*\vXh(\mathbb{Q}_1^T\mathbb{K}_0^{-1} )\vec{q}^T -\frac{1}{4}\vec{q} \mathbb{K}_0^{-1} \mathbb{Q}_0\mathbb{K}_0^{-1} \vec{q} ^T\right] \right\rbrace 
\end{equation}

Substituting $I_3$ in \ref{part_1_2}  we get:

\begin{equation}
\begin{split}
P(\vXh,\vec{M})\propto\exp\left\lbrace\frac{1}{2} \left[ \vXh\mathbb{Q}\vXh^T+2 \vXh \mathbb{K}  \vXh^T  \right]  \right\rbrace * I_4
\end{split}
\label{part_2_2}
\end{equation}

where 

\begin{equation}
\begin{split}
I_4=&\int d \vec{q} \exp \left\lbrace -\frac{1}{2}  \vec{q} \frac{\mathbb{K}_0^{-1} \mathbb{Q}_0\mathbb{K}_0^{-1} }{4}\vec{q} ^T  -\frac{i}{2}\left(\vXh \mathbb{Q}_1^T \mathbb{K}_0^{-1}+2\vec{M} \right) *\vec{q}^T \right\rbrace \\
=& cte *\exp \left\lbrace -\frac{1}{2} \left(\vXh \mathbb{Q}_1^T \mathbb{K}_0^{-1}+2\vec{M}   \right) \mathbb{K}_0\mathbb{Q}_0^{-1}\mathbb{K}_0 \left(\vXh \mathbb{Q}_1^T \mathbb{K}_0^{-1}+2\vec{M}  \right) ^T\right\rbrace 
\end{split}
\label{part_3_2}
\end{equation}
and replacing $I_4$ in \ref{part_2_2} we finally get:

\begin{equation}
\begin{split}
P(\vXh,\vec{M})\propto&\exp\left\lbrace-\frac{1}{2} \left[ -2\vXh\mathbb{K}\vXh^T+4 \vXh \mathbb{Q}_1^T\mathbb{Q}_0^{-1}\mathbb{K}_0 \vec{M}^T  +4*\vec{M} \mathbb{K}_0\mathbb{Q}_0^{-1}\mathbb{K}_0 \vec{M}^T\right] \right\rbrace \\
\propto&\exp\left\lbrace-\frac{1}{2} \vec{z} \mathbb{V}^{-1}\vec{z}^T\right\rbrace 
\end{split}
\end{equation}

with $\vec{z}=\left[ \vXh,\vec{M}\right] $ and 

\begin{equation} 
\mathbb{V}^{-1} =  \left(
\begin{array}{cccc}
{ -2 \mathbb{K}}&{ 2 \mathbb{Q}_1^T\mathbb{Q}_0^{-1}\mathbb{K}_0}\\
{ 2\mathbb{K}_0 \mathbb{Q}_0^{-1}\mathbb{Q}_1}& {4 \mathbb{K}_0\mathbb{Q}_0^{-1}\mathbb{K}_0 }\\\
\end{array}\right)
\end{equation}

inverting $\mathbb{V}^{-1}$ we get:
\begin{equation} 
\mathbb{V} =  \left(
\begin{array}{cccc}
{ -\left[2\mathbb{K}+\mathbb{Q}\right]^{-1} }&{ \frac{1}{2}\left[2\mathbb{K}+\mathbb{Q}\right]^{-1}*\mathbb{Q}_1^T\mathbb{K}_0^{-1}}\\
{\frac{1}{2}\mathbb{K}_0^{-1}\mathbb{Q}_1\left[2\mathbb{K}+\mathbb{Q}\right]^{-1} }& {\frac{1}{4}\mathbb{K}_0^{-1}\left[ \mathbb{Q}_0-\mathbb{Q}_1\left[2\mathbb{K}+\mathbb{Q}\right]^{-1} \mathbb{Q}_1^T\right] \mathbb{K}_0^{-1}}\\\
\end{array}
\right)
\end{equation}

which allows to compute the error as:

\begin{equation} 
\hat{d}(\bm C,\bm C_0)=\sum_i \left( \left[ \mathbb{V}_{11}\right] _{ii}-2*\left[ \mathbb{V}_{12}\right]_{ii}+\left[  \mathbb{V}_{22}\right]_{ii} \right) 
\end{equation}

\subsection{Initializing parameters} 
\label{sub:initializing_parameters}

\subsubsection{Time scale parameter $\mu$ for  FastML inference}
\label{ap:parameter_FastML}
 The mutation rate parameter  $\mu$ in the Felsenstein model \ref{single_site_propagator} is  typically  unknown,  then it must be inferred  from the data. Note that under model \ref{single_site_propagator} the average Hamming distance between two equilibrium sequences at evolutionary time distance $\Delta t$ can be computed as

\begin{equation}
\bar{d}_H(\Delta t)=\left( 1-e^{-\mu \Delta t}\right) \bar{d}_H(\infty)
\label{av_hamming_dist}
\end{equation}

where average $\bar{d}_H(\infty)$ is the Hamming distance between two independent equilibrium sequences in the independent-site model. Therefore, we can take any two sequences at leaves of the tree, calculate their Hamming distance together with their time separation on the phylogenetic tree by adding all branch lengths along their connecting path, and use the result as an instance of $d_H(\Delta t)$. Taking all pairs of sequences from the alignment, we can bin the observed times, calculate average Hamming distances for each time bin, and fit the functional form of equation \ref{av_hamming_dist} to obtain the desired value of $\mu$. 

\begin{figure*}[!htb]
	\begin{center}
		
		\centering\includegraphics[keepaspectratio=true,width=0.5\textwidth]{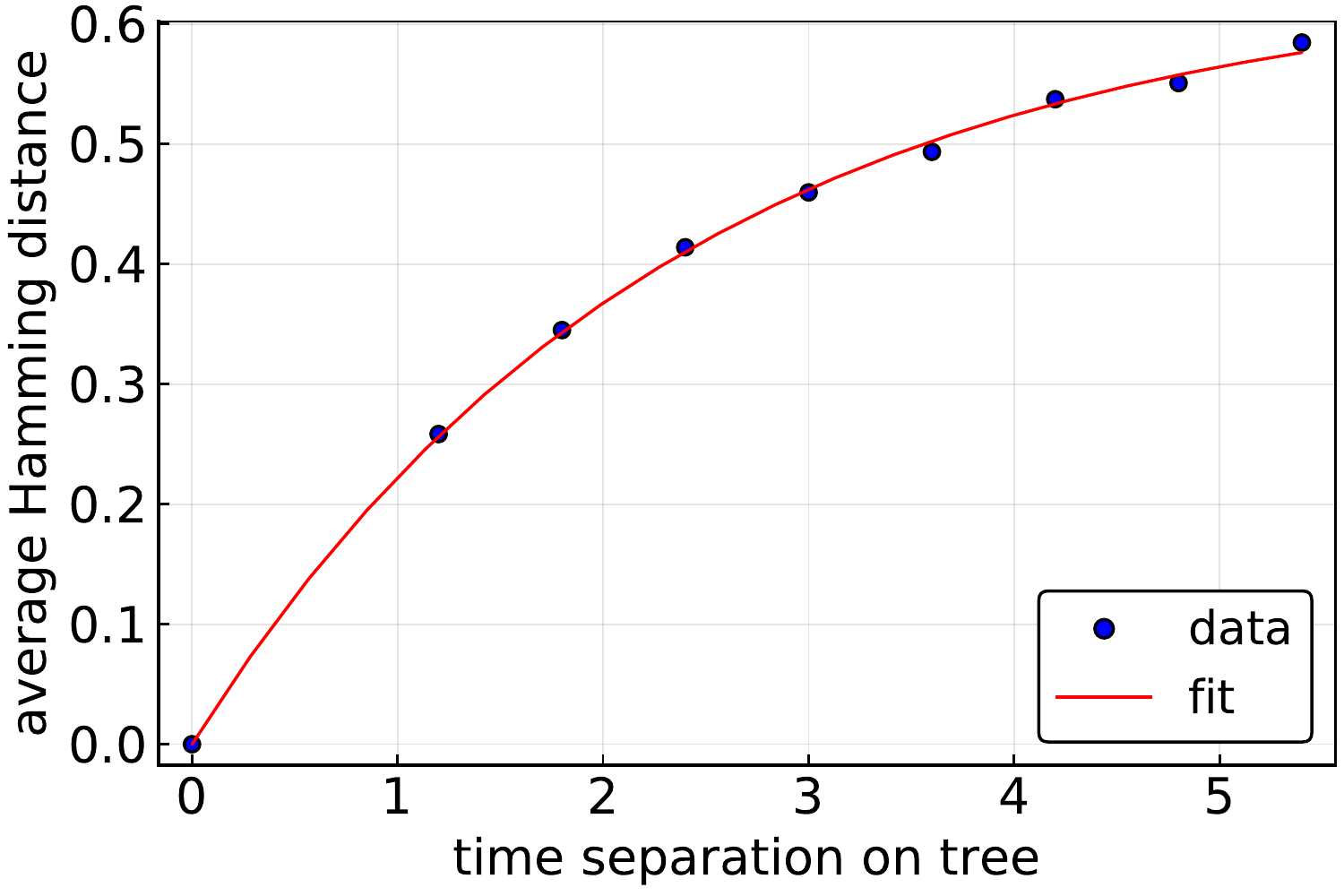}
		
	\end{center}
	\vspace{-1mm}
	\caption{{ Fitting $\mu$ }} \label{fitting_mu}
\end{figure*}

\begin{figure*}[!htb]
	\begin{center}
		\begin{subfigure}
			\centering\includegraphics[keepaspectratio=true,width=0.45\textwidth]{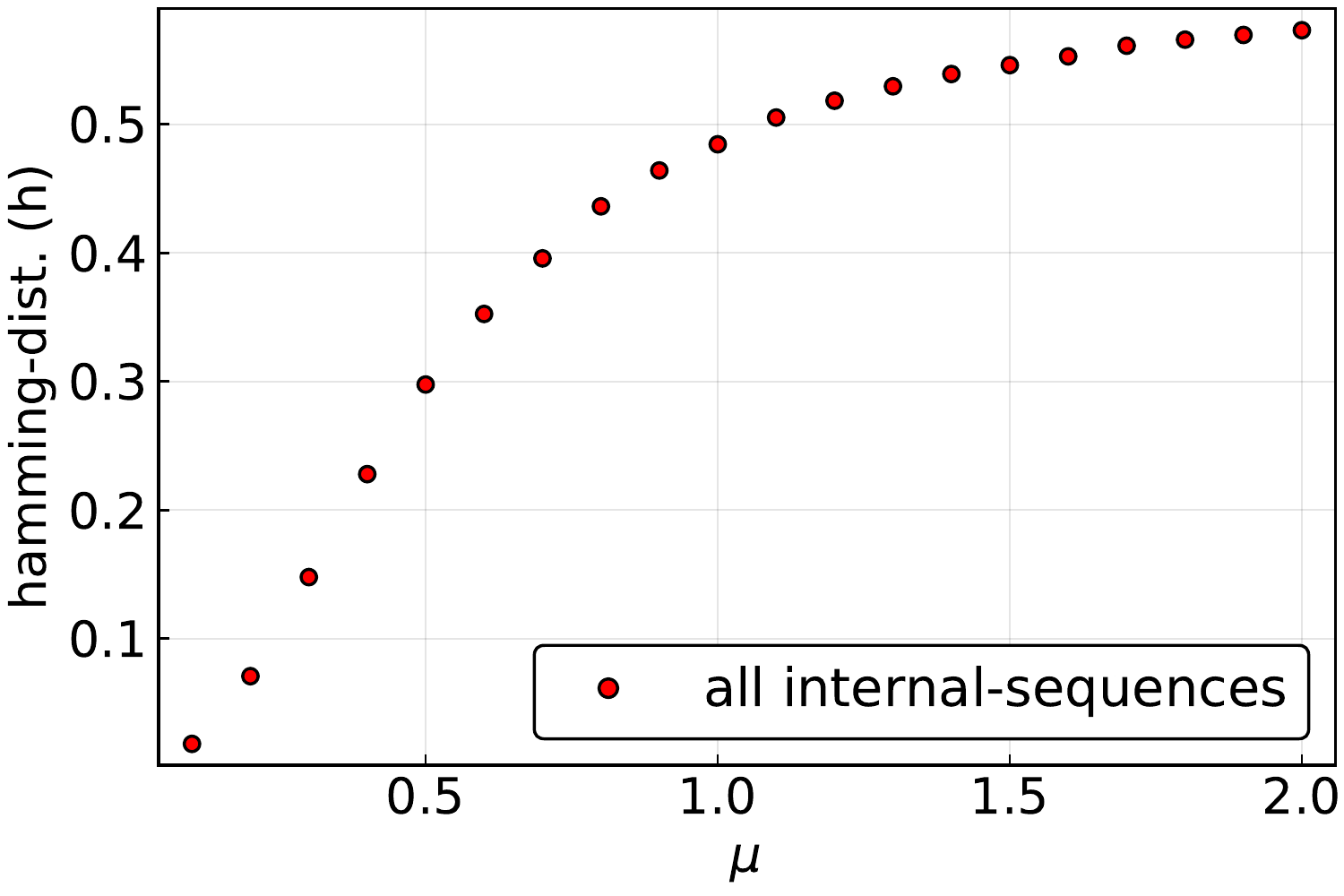}
		\end{subfigure}
		\hspace{1mm}
		\begin{subfigure}
			\centering\includegraphics[keepaspectratio=true,width=0.45\textwidth]{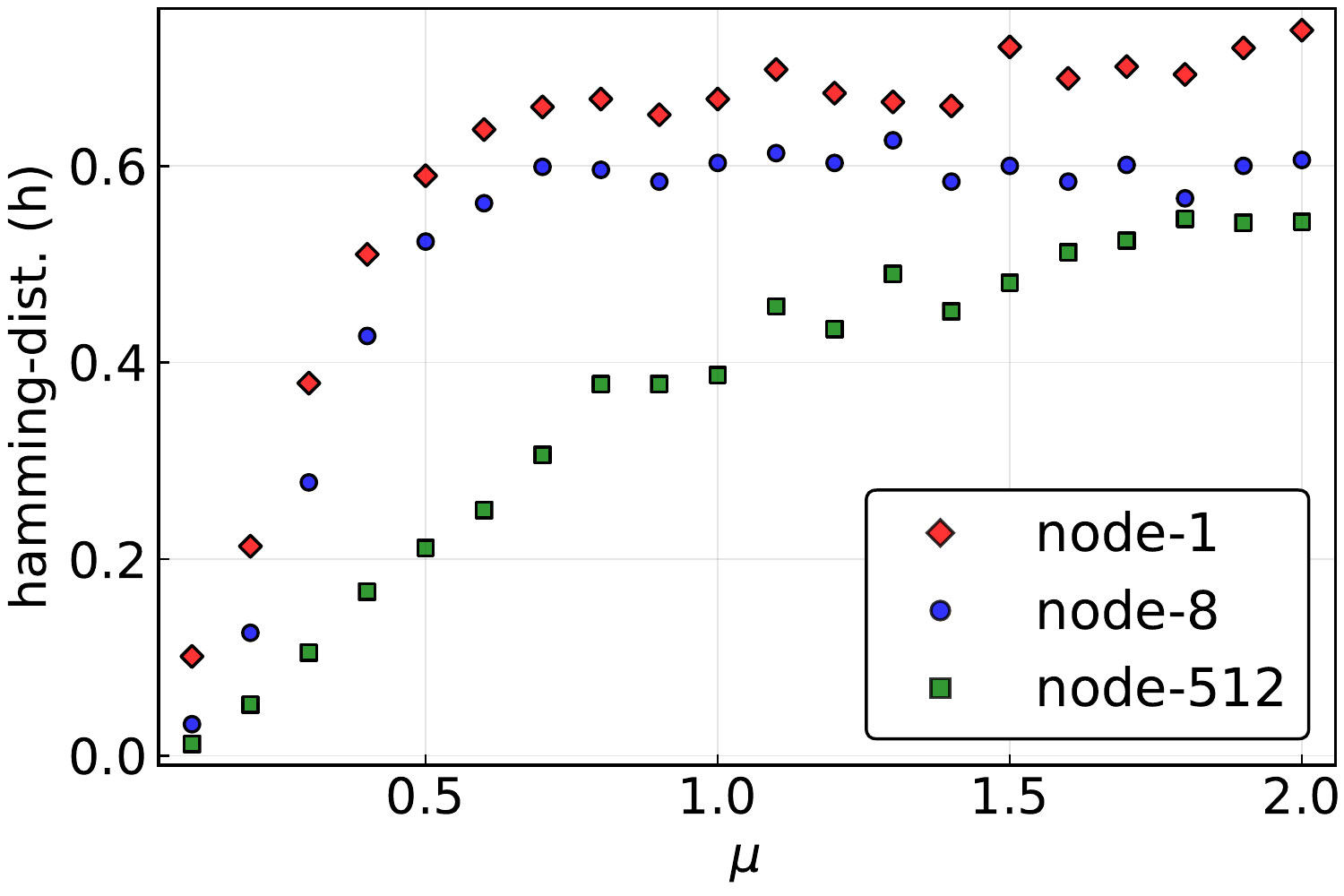}
		\end{subfigure}
	\end{center}
	\vspace{-1mm}
	\caption{{  Ancestral reconstruction via our implementation of FastML algorithm. \textbf{left:} Distance between inferred and true full set of ancestral sequences. \textbf{right:} Distance between  inferred and true sequences at different internal nodes in the tree. }} \label{joint_fastML}
\end{figure*}

As proof of concept, we  show our implementation of this algorithm for  data generated with  single site model \ref{single_site_propagator} on a homogeneous and binary tree with $H=9$ bifurcation events.  Figure \ref{fitting_mu} show results of fitting equation \ref{av_hamming_dist}  and  Figure \ref{joint_fastML} show the distance between inferred and true ancestral sequences  for different values of the mutation rates. From figures, it is possible to note that reconstruction get worse for higher mutations rate and for deepest internal nodes, an expected feature for this problem.

\subsubsection{Time scale parameter $\gamma$}
\label{Inferring gamma}
The timescale parameter $\gamma$ is a priori unknown and must be inferred from data as an initial step before the  ancestral reconstruction algorithm. 
Since the process is always Gaussian, the distribution of the leaves has to be Gaussian itself. Furthermore, we know the covariance between any two elements. Within one leaf, the covariance is the equilibrium covariance $\bm C $, and among leaves it has to be $\Lambda^{\Delta t_{ij}}{\bm C}  $ with $\bm \Lambda = \exp(-\bm C^{-1}*\gamma)$ and $\Delta t_{ij}$ is the  path time   between nodes $i$ and $j$ along the branches of the tree.

Then the  leaves distribution is given by 
\begin{equation}
\nonumber
P(\vXd=\left[ \vec{x}^1_l,\dots,  \vec{x}^{N_{l}}_l \right]  ) = \frac 1 {\sqrt{ 2 \pi^{N_{l}*L} \det \mathbb{G}}  }\times \exp\left(-\frac 1 2 \vXd \mathbb{G}^{-1}\vXd^T \right)
\label{eqn:leaves_dist}
\end{equation}
where $\mathbb{G}$ is a block matrix whose structure is induced by the  phylogenetic tree  and   its elements are given by:

\begin{equation}
\bm G_{i,j} =
\begin{cases}
\bm C & \text{$i=j$}.\\
\bm \Lambda^{\Delta t_{ij}}{\bm C} & \text{otherwise}  
\end{cases}
\end{equation}

Then the  likelihood is given by
\begin{equation}\label{eq:log_likelihood}
L(\vXd|\bm C,\gamma) = \frac 1 2 \log \det  \mathbb{G}^{-1}  -\frac 1 2 \vXd\mathbb{G}^{-1} \vXd^T
\end{equation}

depending only on the tree as well as on $\bm C$ and $\gamma$.  As we know the tree and the covariance matrix $\bm C$ we can obtain the $\gamma$ parameter by maximizing the likelihood function.

\end{document}